\documentclass[11pt]{article} 
\usepackage{amsmath,amsthm,latexsym,amssymb,amsfonts,epsfig}

\newtheorem*{Theorem}{Theorem}
\newtheorem*{Definition}{Definition}

\newcommand{\be}{\begin{equation}}
\newcommand{\ee}{\end{equation}}
\newcommand{\ba}{\begin{eqnarray}}
\newcommand{\ea}{\end{eqnarray}}
\newcommand{\bv}{\bar{v}}

\title{{\sf On propagation in Loop Quantum Gravity}}
\author{
{\sf T. Thiemann}$^1$\thanks{{\sf 
thomas.thiemann@gravity.fau.de}}\\
{\sf M. Varadarajan}$^2$\thanks{{\sf 
madhavanvaradarajan@yahoo.co.in}},
\\
{\sf $^1$ Inst. for Quantum Gravity, FAU Erlangen -- N\"urnberg,}\\
{\sf Staudtstr. 7, 91058 Erlangen, Germany}\\
{\sf $^2$ Raman Research Institute, C. V. Raman Avenue,
Bangalore -- 560 080, India}\\
}
\date{{\small\sf \today}}

\makeatletter
\@addtoreset{equation}{section}
\makeatother

\begin{document} 

\maketitle

{\sf

\begin{abstract}
A rigorous implementation of the Wheeler-Dewitt equations was derived 
in the context of Loop Quantum Gravity (LQG) and was coined Quantum Spin
Dynamics (QSD). 
The Hamiltonian constraint of QSD was criticised as  
being too local and to prevent ``propagation'' in canonical LQG.
That criticism was based on an algorithm developed 
for QSD 
for generating solutions to the Wheeler-DeWitt equations. The 
fine details of that algorithm could not be worked out because the QSD 
Hamiltonian constraint makes crucial use of the volume operator which 
cannot be diagonalised analytically. 

In this paper, we consider the U(1)$^3$ model for Euclidean vacuum LQG 
which consists in replacing the structure
group SU(2) by U(1)$^3$ and otherwise keeps all properties of the SU(2) theory
intact. This enables analytical calculations and the fine details 
of the algorithm {\it can} be worked out.

By considering  one of the simplest possible non-trivial class of solutions 
based on very small graphs, we show that 1. an infinite number 
of solutions {\it exist} which are 2. generically {\it not normalisable} 
with respect to the inner product on the space of spatially diffeomorphism 
invariant distributions and 3. generically display {\it propagation}.

Due to the closeness of the U(1)$^3$ model
to Euclidean LQG, it is extremely likely 
that all three properties hold also in the SU(2)
case and even more so in physical Lorentzian
LQG. These arguments can in principle 
be  made water tight using modern numerical (e.g. ML or QC) methods 
combined with the techniques developed in this paper which we reserve for 
future work.    
\end{abstract}

\section{Introduction}
\label{s0}

One of the most important unsettled research questions in Loop Quantum 
Gravity (LQG) \cite{Books} is the precise implementation of the 
quantum dynamics, i.e. the quantum Einstein equations or Wheeler-DeWitt 
equations. A concrete derivation of such equations has been given 
in \cite{QSDI}. It starts from the classical Hamiltonian constraint,
which is then discretised in terms of non Abelian 
magnetic holonomy and electric flux variables familiar from lattice gauge 
theory \cite{Creutz} which allow to to define a regularised operator 
on the dense domain $\cal D$ of the kinematical Hilbert space ${\cal H}_{{\rm kin}}$ 
consisting of the span of spin network functions \cite{GraphSymm,LOST}.
The corresponding regulator is essentially the spatial extension of 
loops associated to vertices and pairs of adjacent edges of the graph on which 
the operator acts and the corresponding Wilson loop functions is an 
approximant to the curvature that appears in the Hamiltonian constraint.
It is possible to remove the regulator using an operator topology 
(coined URST in \cite{QSDI}) that exploits the fact that a solution 
to all quantum constraints must be spatially diffeomorphism invariant and 
that therefore the subspace of spatially diffeomorphism invariant vectors 
in the algebraic dual ${\cal D}^\ast$ of $\cal D$ is naturally available 
in order to define a kind of weak$^\ast-$topology. An important property 
of the classical 
Hamiltonian constraint is that it is a scalar density of weight one 
which is kept in the quantum theory and grants the associated covariance 
conditions.  

Thus, \cite{QSDI} provides a concrete proposal (modulo the axiom of 
choice) for the Hamiltonian 
constraint of vacuum quantum gravity with Euclidean or Lorentzian 
signature in the continuum, densely defined on $\cal D$. 
It is free of anomalies in the sense that 
the action of its commutators on the algebraic dual 
annihilate spatially diffeomorphism invariant elements. However, 
the Hamiltonian constraint operator constructed in \cite{QSDI} is not 
yet entirely satisfactory for the following reason: While the dual 
of its commutators annihilate spatially diffeomorphism invariant 
distributions and thus must be a linear combination of spatial 
diffeomorphism constraints \cite{GraphSymm}, that linear combination 
does not qualify as the linear combination that one would expect from
a direct quantisation of the corresponding classical 
Poisson bracket \cite{QSDIII} performed by the same methods. Using the 
analogy of Lie algebras (or algebroids), the quantum constraint algebra 
closes but it closes with the wrong structure constants (or functions),
the correct ones representing the hypersurface deformation algebra. 

Correspondingly, several methods for improvement have been suggested 
in the literature. These efforts can be subdivided into two classes:
In the first class, the issue of the structure functions is avoided 
altogether, in the second, the correctness of the structure constants 
is used as a guiding principle to adjust the fine details of the 
construction proposed in \cite{QSDI}. In historical order, 
the first class comprises the 
master constraint approach \cite{Master} and the reduced   
phase space quantisation  approach \cite{KGTT} while the second class 
comprises the electric shift approach \cite{MV} (see also \cite{laddha} for a preliminary attempt)
and the Hamiltonian 
renormalisation approach \cite{TTRen} which can be applied 
to both the constraint of \cite{QSDI} and the physical Hamiltonian of 
\cite{KGTT} (see also the references in all
four manuscripts). Common to the \cite{QSDI, MV} 
versions of the Hamiltonian 
constraint is that the Hamiltonian constraint acts non-trivially only at 
the vertices of the graph and its action on a given vertex 
deforms the graph in an open neighbourhood
of that vertex not modifying the graph in neigbourhoods of the 
other vertices (in \cite{QSDI,MV} this deformation is encoded  
by the loop approximant to the curvature that one uses). By contrast, 
in \cite{Master, KGTT,TTRen} such a deformation is not considered as the 
loops involved are part of the same lattice on which the graph in question
is defined where the lattice plays the role of a resolution scale 
in the sense of the Wilsonian point of view of renormalisation and 
at that scale one computes the matrix elements 
of the Hamiltonmain (constraint) (the continuum operator is then 
some kind of inductive limit of these quadratic forms). 

In that sense, the criticism conveyed in \cite{Smolin} should apply
to the works \cite{QSDI,MV}: The arguments in \cite{Smolin} are based 
on \cite{QSDII} where an algorithm is sketched for how to construct 
solutions in the algebraic dual of both the spatial diffeomorphism
constraint of \cite{GraphSymm} and the Hamiltonian constraint of 
\cite{QSDI}. The construction \cite{QSDII} can be sketched as follows:
A solution of the spatial diffeomorphism constraint is a 
linear combination of the diffeomorphism group averagings 
$\eta(S)$ of spin network
functions (SNWF) $S$
(moulo graph symmetries -- see \cite{GraphSymm} for details) which 
form an orthonormal basis of the kinematical Hilbert space.
Given such an element $\Psi$, one must impose that its evaluation on 
$C(N)S$ vanishes for all $N,S$ 
where $C(N)$ is the Hamiltonian constraint with  
lapse function (smearing function of the Hamiltonian constraint) $N$ and 
$S$ is any spin network function. Since $N$ 
can be chosen to be supported only in the vicinity of a vertex $v$ 
of the graph $\gamma(S)$ on which $S$ is supported and since $C(N)$ acts 
non-trivially only at those vertices (with action $C_v$) 
we have equivalently that 
$\Psi(C_v S)=0$ for all $S$ and all $v\in V(\gamma(S))$ (vertex set
of the graph $\gamma(S)$ underlying $S$). 
Since $C_v$ modifies $\gamma(S)$ and the colourings $m(S)$ of $\gamma(S)$ (by 
spin quantum numbers on edges and intertwiners at vertices) only in 
the vicinity of $v$ leaving the rest of $S$ unmodified 
(modulo diffeomorphisms) the following strategy suggests itself:
$C_v S$ can be decomposed into SNWF $S'$ over new graphs $\gamma_v$.
Due to the details of the graph deformations $\gamma_v$ of \cite{QSDI} which 
adds ``extraordinary edges'', there are graphs $\gamma^{(0)}$ 
which can never be of the form $\gamma_v$ and thus the diffeomorphism 
group averagings of SNWF over such $\gamma^{(0)}$ are exact solutions to 
all constraint equations. We call these solutions ``topological'' 
because they are even normalisable with respect to the 
inner product on the space of solutions to the spatial diffeomorphism 
constraint which is not to be expected for generic solutions. To 
construct more interesting solutions, we are thus led to consider 
(diffeomorphism averages) of graphs which are of the form 
$\gamma^{(1)}=\gamma^{(0)}_v,\;v\in V(\gamma^{(0)})$, from those graphs
we generate graphs 
of the form $\gamma^{(2)}=\gamma^{(1)}_v, v\in V(\gamma^{(1)})$ 
and proceeding inductively we generate 
graphs of the form $\gamma^{(n)}$. We may develop genealogical language 
and call graphs of the respective form $\gamma^{(0)}$ primordial, 
$\gamma^{(n)},\;n>0$ descendants of generation $n$, 
$\gamma^{(n+1)}$ children of the parents $\gamma^{(n)}$. To classify solutions 
of the Hamiltonian constraint it is therefore useful to keep track
of the generation $n$ of a (diffeomorphism average of a) graph descending
from a primordial one.
Similarly we have a genealogy of SNWF which is different from that for 
graphs because also the colourings are involved. No confusion arises 
if we explicitly say child graph or child SNWF etc.

A particular 
feature of the operator \cite{QSDI} is that the action of $C_v$ on SNWF
over 
$\gamma_{v'},\; v'\in V(\gamma)$ is trivial for 
$v\in V(\gamma_{v'})-V(\gamma)$ and for $v\in V(\gamma)$ 
equals the action of $C_v$ on SNWF over 
$\gamma$ (modulo diffeomorphisms).
It is precisely this feature 
which makes the action of $C(N)$ anomaly free. Thus, modulo diffeomorphisms,
all one needs to know in order to let the Hamiltonian constraint act on a
SNWF, when evaluated on a diffeomorphism invariant distribution, 
is a neigbourhood of $S$ (graph and colourings) of each vertex 
which modulo diffeomorphisms can be chosen arbitrarily ``small''. For the 
same reason, it is irrelevant that the axiom of choice is involved in the 
definition of $C$ because different choices are washed away by the 
diffeomorphism averaging and the diffeomorphism invariant characteristics
do not need the axiom of choice.

As shown in \cite{QSDII}, the space of solutions to all constraints 
for the Euclidean version acquires the following 
structure\footnote{Note that the arXiv version of \cite{QSDII} 
differs from its published version. Both versions discuss non-symmetric 
and symmetric operators but these operators are different 
in the two versions. What follows applies to the non-symmetric operator 
in the arXiv version, which coincides with the operator 
of \cite{QSDI}, which the discussion in \cite{Smolin} refers to.  
While surprisingly non-symmetric operators are the ones of physical interest 
due to the arguments given in \cite{Kuchar},   
nevertheless we will also comment on a symmetric 
operator which however is different from the one discussed in the 
arXiv version of \cite{QSDII}, see section \ref{s1.3a}. Both $U(1)^3$ analogs 
of the non-symmetric and the symmetric operator discussed in this 
paper display propagation.}: \\
i. It is a linear combination of solutions of generation $n$ for 
$n=0,1,2,..$.
Here a solution of generation $n$ is a linear combination of diffeomorphism
averages of SNWF of generation $n$.\\
ii. Such a solution of 
generation $n$ is itself a linear combination
of elementary solutions of generation $n$. An elementary solution 
of generation $n$ is a linear combination of of diffeomorphism 
averages of SNWF which are n-th generation children of SNWF with the same 
primordial parent graph\footnote{\label{fn2}This property holds only for the 
published version of \cite{QSDII}. In the arXiv version, contrary to what 
is assumed there, in fact disappearance of edges in child graphs 
is possible so that there is not a single primordial graph leading 
to a given child graph. In this paper, as we use the non symmetric
version of the Hamiltonian constraint \cite{QSDI},
we take this effect into account when we construct our solutions. We discuss these matters further in 
section \ref{s1.3a}}.\\
For the Lorentzian version property i. does not hold as the 
solutions cannot be built from elementary solutions of constant 
generation $n$, necessarily more generations are involved and 
it cannot be excluded that a
non-trivial typical solution is a linear 
combination of diffeomorphism averages of SNWF of unbounded 
generation. On the other hand, property ii. still holds, elementary 
solutions are built from the same primordial graph. 
  
Now assume that in addition also the following property holds 
\cite{Smolin} (irrespective 
of considering the Euclidean or Lorentzian version):\\
iii. For an elementary solution $\Psi=\sum_{l\in L}\; \kappa_l\; l$
where $l\in L$ is of the form $\eta(S)$ and $\gamma(S)$ is some descendant
of the same primordial graph $\gamma^{(0)}$, the label set $L$ can be 
partitioned into susbsets $L_v$ with    
$v\in V(\gamma^{(0)})$ such that the constraint equations resulting from 
$C_v$ have non-trivial influence only on $\kappa_l$ for $l\in L_v$. \\
If this property, which due to the local action of $C_v$ appears 
to be obvious, would hold then a local perturbation of the solution
i.e. a variation of the $\kappa_l$ for $l\in L_v$ has no influence 
on the $\kappa_{l'},\; l'\in L_{v'},\; v'\not= v$. 
That is \cite{Smolin}, the conclusion 
would be that there is {\it no propagation}
or {\it absence of long range correlations}
in the space of solutions to the Hamiltonian constraint of \cite{QSDI} 
in the sense that the structure 
of the solution at vertex $v$ is local to $v$ and has no effect at any other 
vertex $v'$. Similar remarks would hold for the Hamiltonian constraint 
of \cite{MV}.\\ 
\\

In what follows we focus on the property of propagation in the context of the  
constraint action of \cite{QSDI}. With regard to propagation in the context of the 
constraint action of \cite{MV}, we restrict ourselves to a few remarks in section \ref{s1.3a},
these remarks being indicative of a reasonable expectation that this constraint action
does display propagation. A detailed analysis of propagation in the context of the 
constraint \cite{MV} is left for future work.

Accordingly, the main purpose of the present paper is twofold: \\
I. First, we show that assumption iii. above is generically violated in LQG, 
independent of the signature.
In fact this was implicitly
known since the appearance of \cite{QSDI}. \\ 
II. Second, even if assumption iii. is violated, the conclusion stated 
in \cite{Smolin} may still hold as assumption iii. is not obviously 
a necessary 
condition to hold for the absence of propagation to occur. However, 
we will show that also the conclusion is 
{\it extremely likely to be false} in LQG independent of the 
signature.\\
In that sense, propagation in 
QSD was present ever since it was proposed.\\
\\ 
Let us explain this in more detail:

The mechanism behind the violation of condition iii.
is the diffeomorphism averaging that occurs in the solution to all 
constraints which prohibits a partition labelled by vertices of the 
underlying primordial graph as described above 
and leads to {\it non-unique parentage} at the 
level of diffeomorphism averages of SNWF. The fact, that non-unique 
parentage is the {\em key} ingredient for propagation in the context of
LQG constraint quantization methods,  was first realized in the toy model context of 
Parameterized Field Theory \cite{proppft}.
In fact, the mechanism of non-unique parentage can also seen to be
responsible for the absence of anomalies \cite{QSDI} in the sense described 
above and does not rely on the details of the action of the Hamiltonian
constraint.

Still, it may accidently happen that long range correlations are absent
in solutions to the Hamiltonian constraint and it is here where details 
matter. The reason why in contrast to I.
we cannot make a statement with certainty in II,  
is due to the fact that the volume operator is 
a central ingredient of the operators \cite{QSDI} whose spectrum 
cannot be computed analytically. Accordingly, precise computations
and estimates are not possible, thus prohibiting explicit solutions.
To make progress, in this paper we consider the $U(1)^3$ model of 
Euclidean quantum gravity invented in \cite{Smolin2} which shares 
almost all of its features except that the non-Abelian group 
SU(2) is replaced by U(1)$^3$. We do this for the sole purpose of
being able to compute the corresponding volume spectrum analytically.
We are then able, for sufficiently simple 
primordial graphs, to solve the Hamiltonian constraint equations 
analytically 
thus allowing to compute the linear combinations 
of diffeomorphism invariant distributions that define the solutions explicitly
together with the complete parametrisation of the freedom involved.
This parametrisation freedom manifests the presence of long range correlations 
or propagation. 
While the explicit calculations are confined to a particularly simple 
primordial graph, it follows from the solution generating technique that 
we develop that propagation happens also for higher generations 
and for arbitrarily long iterations 
and gluings of those simple graphs filling all of a Cauchy surface arbitrarily 
densely. It will also become clear from the example how to compute the 
solutions explicitly in the general case.
These strong statements can so far not be made within the 
SU(2) theory due to the complexity of the volume spectrum. However, 
and this is the sense in which we use the term ``extremely likely'' above,
it is beyond reasonable doubt that the spectrum of the SU(2) volume operator 
does not conspire in such a way as to render the conclusions derived for 
the U(1)$^3$ model invalid. Since the Lorentzian Hamiltonian constraint
uses the Euclidean version and multiple commutators thereof with 
the volume operator \cite{QSDI}, the same conclusion 
is  extremely likely to hold for Lorentzian LQG. 

As a byproduct of our computations 
we are also able to show that an infinite number of non-trivial solutions 
{\it exist} and that the generic solution to the Hamiltonain 
constraint {\it is not normalisable} with respect to the inner product 
on the space of spatially diffeomorphism invariant distributions 
\cite{GraphSymm}. The possibility of non-existence of non-trivial 
(i.e. non-topological) or absence of non-normalisable solutions 
can again not be ruled out for 
the SU(2) theory of \cite{QSDII} with certainty but is again 
beyond reasonable doubt.

The lesson learnt from the present work is that, modulo the reservation
just spelled out, at least with respect to 
the existence, normalisability and propagation aspect 
the constraint operator \cite{QSDI} does not suffer from 
pathologies and we expect 
that modifications thereof that faithfully implement the hypersurface 
algebra will share this property.\\
\\
The architecture of this article is as follows:\\
\\

In section two we recall the essential ingredients of LQG necessary for the 
understanding of the present work, sketch the structure of the solutions 
of the Hamiltonian constraint and define the notion of propagation desired 
of solutions of the Hamiltonian constraint in LQG.

In section three we define the simple primordial graph and its 
descendants of generation one for which we perform explicit calculations
in the U(1)$^3$ model of Euclidan LQG. We establish 
{\it existence, non-normalisability and presence of long range 
correlations} in its solution space. 

In section four we conclude and give an outlook into further 
study. We state in more detail why these features 
extend to generic arbitrarily large graphs and extremely likely to
the SU(2) case in both Euclidean and Lorentzian signatures.
In particular we encourage 
the application of numerical methods in LQG \cite{QGC, Num} perhaps combined with 
modern machine learning \cite{AI,TN} or 
quantum computing \cite{QC}
techniques in order to perform an actual SU(2) calculation 
involving the 
SU(2) volume operator 
at least numerically using the explicitly known matrix elements 
of its fourth power \cite{JBTT} in order to complete the rigorous 
proof in the SU(2) theory.

\section{LQG, Quantum Einstein Equations and Propagation}
\label{s2}

We begin by recalling the essential notions from LQG and the corresponding 
U(1)$^3$ model. See \cite{QSDI} for details on the Hamiltonian constraint,
reference \cite{TTRen} for a recent 
low-technical review of LQG and \cite{MV1} for more details on the U(1)$^3$ 
model invented in \cite{Smolin2}. 
After that we review the solution generating algorithm of 
\cite{QSDII} and the notion of propagation defined in \cite{Smolin} slightly
adapted to our language.

\subsection{Essentials from LQG}
\label{s2.1}

A closed graph $\gamma$ is the union of its semi-analytic 
\cite{LOST} oriented edges $e\in E(\gamma)$ that intersect nowhere except
in its endpoints which are called vertices $v\in V(\gamma)$. Each 
of its edges $e$ carries a colouring by an irreducible representation   
of the underlying gauge group G which is assumed to be compact, that is,
a half-integral spin quantum number $j_e$ in case of G=SU(2) and 
a vector $m_e\in\mathbb{Z}^3$ in case of G=U(1)$^3$. 
We are interested in solutions to all constraints and thus 
for reasons of Yang-Mills type of 
gauge invariance, each vertex 
$v$ is coloured by an intertwiner that intertwines the tensor 
product of the represesentations 
labelling the outgoing edges and the contragredients of the representations 
labelling the incoming edges with the trivial represention. In case of 
SU(2) the intertwiners $\iota_v$ are essentially invariant tensors built from 
Clebsch-Gordon coefficients, in case of U(1)$^3$ it is simply a Kronecker 
symbol $\delta_v$ 
that imposes that the sum of incoming charge vectors equals the sum of 
outgoing charge 
vectors. A spin network function (SNWF) S is a function of the underlying 
connection $A$ which is constructed from the corresponding holonomies 
$A(e)$ along the edges $e$ by plugging them into the matrix element 
functions of the corresponding irreducible representation and contracting 
their matrix element indices with the intertwiner indices. They are 
normalised with respect to the product Haar measure where the number
of factors euqals the number of edges. The kinematical Hilbert space 
${\cal H}_{{\rm kin}}$ 
can be defined by declaring the SNWF to be an orthonormal basis. For 
SU(2) we would label $S=T_{\gamma,j,\iota}$ where $j,\iota$ stands for 
the collection of spins and intertwiners respectively, for U(1)$^3$ we 
write $S=T_{\gamma,m}$ dropping the trivial intertwiner label and assuming 
that the constraints on the collection of charges $m$ hold. To have 
a unified notation, we introduce the compound label $m=(j,\iota)$
also in the SU(2) case. The span 
${\cal D}$ of SNWF defines a dense and invariant domain for the Hamiltonian
constraint \cite{QSDI}. We define $\gamma(S)=\gamma,m(S)=m$

The group of spatial diffeomorphisms $\varphi\in$Diff($\sigma$) of the model 
Cauchy surface $\sigma$ underlying the canonical formulation of LQG 
is represented unitarily on ${\cal H}_{{\rm kin}}$ and $U(\varphi)$ 
acts on SNWF by dragging $\gamma$ and its colourings along, i.e.
$\gamma\mapsto \varphi(\gamma),\;j_{\varphi(e)}=j_e,\;\iota_{\varphi(v)}  
=\iota_v$ for SU(2) and $m_{\varphi(e)}=m_e$ for U(1)$^3$. 
With some care \cite{GraphSymm}, 
one can ``average'' SNWF over Diff$(\sigma)$ resulting in distributions 
$\eta(S)$ on ${\cal D}$, that is elements of the algebraic dual 
${\cal D}^\ast$ of 
$\cal D$ (linear functionals on ${\cal D}$ without continuity conditions).
In case that $\gamma(S)$ has no graph symmetries (i.e. diffeomorphisms 
that preserve $\gamma$ as a set but permute the edges) the linear 
functional $\eta(S)$ is defined by $\eta(S)[S']$ to equal unity when
$S,S'$ differ by a diffeomorphism and zero otherwise (we will only need 
$S$ without graph symmetries in this paper). One parameter groups
of Diff$(\sigma)$ do not act strongly continuously on the kinematical Hilber
space so that there is no spatial diffeomorphism constraint operator, but
one may still impose the spatial diffeomorphism constraint on 
$\Psi\in{\cal D}^\ast$ in the form 
$\Psi[(U(\varphi)-{\rm id}_{{\cal H}_{{\rm kin}}})S]=0$ for all $S$ and 
$\varphi$. Thus the general solution to those constraints are linear 
combinations of the $\eta(S)$ and their linear span can be equipped with the 
inner product $<\eta(S),\eta(S')>_{{\rm Diff}}=\eta(S')[S]$ (again modulo 
details associated with graph symmetries) with respect to which the 
$\eta[S]$ form an (almost) orthonormal basis. 

The Euclidean Hamiltonian constraint $C(N)$ of \cite{QSDI} for smearing (lapse) 
function $N$ is densely defined on ${\cal D}$ and 
for SU(2) and U(1)$^3$ has the following general form
\be \label{2.1}
C(N)\; T_{\gamma,m}=\sum_{v\in V(\gamma)}\; N(v)\; C_{\gamma,v}\;
T_{\gamma,m},\;\;
C_{\gamma,v}=\sum_{e,e'\in E(\gamma);e\cap e'=v}\; C_{\gamma,v,e,e'}
\ee
We will spell out more details 
about $C_{\gamma,v,e,e'}$ in the next section. For the current section,
it is sufficient to mention that $C_{\gamma,v,e,e'}$ depends on the 
volume operator localised at $v$, holonomies along partial segments 
$s_{\gamma,v,e}$ outgoing from $v$ of edges $e$ adjacent to $v$ 
and arcs $a_{\gamma,v,e,e'}$ between the end points of 
$s_{\gamma,v,e}, s_{\gamma,v,e'}$ which do not intersect $\gamma$ anywhere 
else. The arcs form the extraordinary structure of a graph and a graph 
without extraordinary edges is called primordial. The endpoints of 
extraordinary edges are called extraordinary vertices which are co-planar
but not co-linear and tri-valent. 
Repeated actions of the Hamiltonian constraint do not act at the extraordinary 
vertices but create extraordinary edges ever closer to the vertices of the 
primordial graph. Besides adding extraordinary edges, $C_{\gamma,v,e,e,'}$ 
changes the colourings on $s_{\gamma,v,e},\; s_{\gamma,v,e'},\; v$. 
Graphs or SNWF with
$n$ extraordinary edges that arise in the $n-th$ action of the Hamiltonian 
action arising from SNWF over a primordial graph are called of 
n-th generation.

\subsection{Solutions of the Quantum Einstein Equations and Propagation}
\label{s2.2}
 
By definition, a physical state $\Psi$ is an element of ${\cal D}^\ast$
which solves besides the spatial diffeomorphism constraint also 
$\Psi[C(N)S]=0$ for all $S$ and $N$. By choosing lapse functions of 
compact support, equivalently 
$\Psi[C_{\gamma,v}\; T_{\gamma,m}]=0$ for all 
$\gamma,m,v\in V(\gamma)$. We consider a linear combination $\Psi$ 
of the $\eta(S)$ with complex 
coefficients $\kappa_{[S]}$ where $[S]$ denotes the 
diffeomorphism orbit of $S$. It is clear that every $\eta(S)$ with 
$\gamma(S)$ primordial is an exact (we call it {\it topological})
solution as $C_{\gamma,v}$ increases the  
number of extraordinary edges by one. Next we consider linear combinations 
of $\eta(S)$ with $\gamma(S)$ having precisely one extraordinary edge 
i.e. graphs in the first generation. The condition on the complex 
coefficients are now only non-trivial when $\gamma$ in 
$\Psi[C_{\gamma,v} T_{\gamma,m}]$ is primordial. We try to 
construct a simple subset of all those solutions 
by a picking a single primordial graph $\gamma^{(0)}$ and all its 
first generation descendants $\gamma^{(1)}$ and all possible colourings 
of those forming SNWF $S'$. Denote by $\Sigma([\gamma^{(0)}])$ all
$[S']$ for these $S'$ and consider 
$\Psi=\sum_{[S']\in \Sigma([\gamma_0])}\;\kappa_{[S']} \; \eta(S')$. 
Then the constraint equation reduces to 
$\Psi[C_{\gamma^{(0)},v}\;T_{\gamma^{(0)},m}]=0$ for
all $m,v\in V(\gamma^{(0)})$ which is still an infinite number of conditions.
Similar simplifications can be achieved for solutions of 
higher generations and also for the Lorentzian constraint, see 
\cite{QSDII} for details.

Typically the number of coefficients $\kappa_{[S']}$ 
exceeds the number of $S=(\gamma^{(0)},m)$ for which the 
constraint equations are not automatically satisfied, at least
in a naive counting (both numbers are infinite), simply because 
the number of children graphs together with all its colourings exceeds 
the number of parent graphs (in this case only one) together with with all 
colourings. Hence one expects a rich number of non-trivial 
solutions. In in the SU(2) theory 
this cannot be granted with certainty because the numbers 
$\eta(S')[C_{\gamma,v,e,e'} T_{\gamma,m}]$ are not analytically 
available and the naive counting involving the subtraction 
of infinities is a dangerous enterprise. Thus even the question of 
{\it existence} 
is not entirely trivial to answer in the SU(2) theory. 
Next, it may or may not be true that there exist 
solutions with only a finite number of non-vanishing coefficients 
$\kappa_{[S']}$ (besides the topological solutions)
which would be normalisable with respect to the diffeomorphism invariant
inner product. If such solutions would exist we would call them {\it bounded}
as they have finite norm with respect to the diffeomorphism invariant 
inner product. On physical grounds, one would expect the interesting 
solutions to be unbounded as the Hamiltonian constraint at constant 
lapse is a diffeomorphism invariant unbounded 
operator on the diffeomorphism invariant Hilbert space so that zero should 
not be exclusively in its point spectrum. Finally we come to the central 
question of the present work:\\
\\
For simplicity we consider the problem of constructing solutions involving 
only diffeomorphism averages of graphs with one extraordinary edge (first 
generation children). Similar remarks hold for solutions involving
only diffeomorphism averages of constant generation $n$ \cite{QSDII}. 
To further simplify the analysis, we may focus on children graphs 
$\gamma^{(1)}\in \Gamma(\gamma^{(0)})$ which result from a single 
primordial parent graph 
$\gamma^{(0)}$ via the action of the Hamiltonian constraint. As the action 
of the Hamiltonian constraint is local to the vertices, 
$\Gamma(\gamma^{(0)})$ is the disjoint union of subsets 
$\Gamma_v(\gamma^{(0)}),\; v\in V(\gamma^{(0)})$ where $\gamma_v^{(1)}\in
\Gamma_v(\gamma^{(0)})$ is a graph label that appears in the SNWF 
decomposition of the vectors $C_{\gamma^{(0)},v}\; T_{\gamma^{(0)},m}$
for some $m$. 
Let us denote for a graph 
$\gamma$ by $M(\gamma)$ the set of its possible colourings $m$. Then 
the Ansatz for candidate solution reads in more detail
\be \label{2.2}   
\Psi=\sum_{v\in V(\gamma^{(0)}}\; 
\sum_{\gamma_v^{(1)}\in
\Gamma_v(\gamma^{(0)})}\;\sum_{m_v\in M(\gamma_v^{(1)})}\;
\kappa_{v,\gamma_v^{(1)},m_v}\; \eta(T_{\gamma_v^{(1)},m_v})
\ee
which involves a countably infinite sum\footnote{Strictly 
speaking only if the graphs in question have no moduli \cite{GraphSymm}
or if we work in a single superselected sector. The graphs that we 
use in concrete calculations in the next section do not have moduli because
its vertices are either 4-valent or 6-valent but with only four 
distinct tangent directions of the adjacent edges.}  
of diffeomorphism invariant 
distributions. Then the quantum Einstein equations reduce to the equations 
\be \label{2.3}
\Psi[C_{\gamma^{0},v} \; T_{\gamma^{(0)},m}]=0\;\;\forall \;\;
v\in V(\gamma^{(0)}),\;m\in M(\gamma^{(0)})
\ee
because for any SNWF over a graph not in the diffeomorphism class 
$[\gamma^{(0)}]$ the equation is identically satisfied and for any 
graph diffeomorphic to $\gamma^{(0)}$ the equations are strictly identical.
Suppose now that  
\be \label{2.4}
\eta(T_{\gamma_v^{(1)},m_v})[C_{\gamma^{(0)},v'}\; T_{\gamma^{(0)},m}]
\propto \delta_{v,v'}
\ee
then (\ref{2.3}) splits into the $|V(\gamma^{(0)})|$ independent 
sets of equations 
\be \label{2.5}
\sum_{\gamma_v^{(1)}\in
\Gamma_v(\gamma^{(0)})}\;\sum_{m_v\in M(\gamma_v^{(1)})}\;
\kappa_{v,\gamma_v^{(1)},m_v}\; 
\eta(T_{\gamma_v^{(1)},m_v})[C_{\gamma^{(0)},v'}\; T_{\gamma^{(0)},m}]=0
\ee
which only involves the coefficients $\kappa_{v,\gamma^{(1)}_v,m_v}$. That
is, the assumption (\ref{2.4}) leads to a decoupling of the system 
(\ref{2.3}) with respect to the vertex label and the sets 
of equations (\ref{2.5}) 
can be solved independently. In particular, setting 
$\kappa_{v',\gamma_{v'}^{(1)},m_{v'}}=0$ for all $v'\not=v_0$ and solving 
(\ref{2.5}) for  $v=v_0\in V(\gamma^{(0)})$ would yield a solution. 
Assuming that non-trivial solutions exist, the solution coefficients 
$\kappa_{v,\gamma_v^{(0)},m_v}$ will involve (typically infinitely many) 
free parameters which 
we collectively denote by $\alpha_v$ and which parametrise the kernel 
of (\ref{2.5}). These $\alpha_v$ therefore correspond to 
{\it observables}\footnote{The exchange operators defined in \cite{Smolin}
are (linear) maps of these $\alpha_v$ and since these map solutions 
to solutions, they provide Dirac observables.},
i.e. gauge invariants which are unconstrained and invariant by the constraints
and their gauge motions and which keep some degree of locality 
as they are associated with the vertex $v$ in the diffeomorphism 
class of $\gamma^{(0)}$. Let us denote by 
$\{\Psi(\{\alpha_v\}_{v\in V(\gamma^{(0)})})\}$ the complete space of solutions 
so obtained. Then perturbing $\Psi(\{\alpha_v\})$ with respect to 
$\alpha_v$ has no influence on any other $\alpha_{v'},\; v'\not=v$.
Since the $\alpha_v$ parametrise solutions of the quantum Einstein equations,
they parametrise, likely very indirectly, histories of spacetime metrics 
on the manifold $\mathbb{R}\times \sigma$. Then the discussion suggests
that performing a spatially local perturbation of a quantum solution
has no global effect. If this picture is correct, then, in the 
Lorentzian theory, we would conclude 
that the quantum solutions are incompatible with the classical solutions
of Einstein's equations which are known to be well posed 
in globally hyperbolic spacetimes which are the spacetimes considered
in canonical quantum gravity 
(the perturbation
of an entire history does not have compact support and thus has an unbounded 
domain of dependence). In particular, thinking of the $\alpha_v$ 
as relational observables that depend on an intrinsic physical 
time parameter $\tau$, a variation of $\tau$ should affect all $\alpha_v$.
While these remarks are not entirely conclusive, the absence of propagation
or long range correlations that follows from (\ref{2.4}) is at least 
worrysome.
 
This potential problem spelled out in \cite{Smolin} rests on the assumption 
(\ref{2.4}) which seems to be quite reasonable as the Hamiltonian constraint
acts arbitrarily closely to $v$ modulo diffeomorphisms. The catch is that 
the notion of closeness becomes void after diffeomorphism averaging.
Thus it may happen that while 
$\gamma^{(1)}_v\in \Gamma_v(\gamma^{(0)})$  and 
$\gamma^{(1)}_{v'}\in \Gamma_{v'}(\gamma^{(0)})$ are different graphs 
$\gamma^{(1)}_v\not=\gamma^{(1)}_{v'}$ for $v\not=v'$ and the 
corresponding SNWF are orthogonal, still
$[\gamma^{(1)}_v]=[\gamma^{(1)}_{v'}]$! The atomic prime example for this 
effect is a graph $\gamma^{(0)}$ with two vertices $v,v'$ 
and four non-coplanar
edges between them (to have non-vanishing volume). Pick two of its edges 
$e,e'$. Then there are contributions from $C_{\gamma^{(0)},v,e,e'}$ and 
$C_{\gamma^{(0)},v',e,e'}$ that attach arcs
$a_{\gamma^{(0)},v,e,e'}$ and 
$a_{\gamma^{(0)},v',e,e'}$ respectively to $\gamma^{(0)}$ defining one of the 
possible $\gamma_v^{(1)},\gamma_{v'}^{(1)}$ respectively. But 
$[\gamma_v^{(1)}]=[\gamma_{v'}^{(1)}]$! Thus, while 
the arcs labelled by $v,v'$ are very ``close'' to $v,v'$ respectively, 
they are also very close to $v',v$ modulo diffeomorphisms. 
Therefore, we may find $T_{\gamma^{(0)},m},T_{\gamma^{(0)},m'}$ with 
$m\not= m'$ but such that the SNWF decompositions of  
$C_{\gamma^{(0)},v}\; T_{\gamma^{(0)},m}$ 
and $C_{\gamma^{(0)},v'}\; T_{\gamma^{(0)},m'}$ 
contains SNWF $S,S'$ with $[S]=[S']$ although 
$<S,S'>_{{\cal H}_{{\rm kin}}}=0$.  
Abusing the notation, we call this 
effect {\it non-unique parentage} of SNWF where the abuse refers to the fact that not $S=S'$ but only
$[S]=[S']$. The effect is triggered by the mechanism of diffeomorphism 
averaging. Indeed, the same mechanism is responsible for the fact that 
the algebra 
of Hamiltonian constraints closes modulo diffeomorphisms \cite{QSDI}. 
Therefore, assumption (\ref{2.4}) 
is {\it false} whenever there is non-unique parentage. Whenever there 
is non-unique parentage, even the Ansatz (\ref{2.2}) is strictly 
speaking incorrect
as there is an over-counting involved: Since some of the 
$\eta(T_{\gamma^{(1)}_v,m_v}), \;\eta(T_{\gamma^{(1)}_{v'},m'_{v'}})$ 
for different $v,v'$ and certain $m_v\in M(\gamma^{(1)}_v),\;m'_{v'}\in
M(\gamma^{(1)}_{v'})$ are in fact identical, the separate coefficients 
$\kappa_{v,\gamma^{(1)}_v,m_v},\;
\kappa_{v',\gamma^{(1)}_{v'},m'_{v'}}$ collapse to a single coefficient
$\kappa_{v,\gamma^{(1)}_v,m_v}+\kappa_{v',\gamma^{(1)}_{v'},m'_{v'}}$.
This fact is the technical reason 
for the {\it coupling} between the constraint 
equations resulting from different $v,v'$. Note that non-unique 
parentage implying the failure of (\ref{2.4}) is a generic feature of 
LQG and does not depend on the fine details of the quantum dynamics. \\
\\
We finish this section by providing a concrete technical definition for 
the presence of propagation in a single first generation 
solution to the Hamiltonian constraint which is motivated by the above 
discussion, \cite{Smolin} and the further analysis in this paper.  
Its virtue is that it is free of any details of how one actually 
finds solutions and thus can be stated rather non-technically. 
A second more technical definition of propagation in a whole class 
of solutions which is closer to the actual construction 
of solutions will be deferred to section \ref{s1.7} after we have 
illustrated the construction algorithm.
Both definitions are
to be considered as a working definitions that may have to be refined in the 
future 
as we gain more experience with propagation in ever more complicated solution
classes.

\subsection{\label{s2.3} Working Definition of Propagation}

\subsubsection*{ Preliminaries} 
A candidate first generation solution $\Psi$ of the quantum 
Einstein equations based on a primordial parent graph $\gamma^{(0)}$  
can be written as 
a linear combination with non-redundant coefficients $\kappa_{{\bar v},l}$
where $\bv$ runs through a subset $V$ of $V(\gamma^{(0)})$ and at 
given $\bv$, $l$ runs 
through a subset $L_{\bv}$ of pairs $\gamma^{(1)}_{\bv},m_{\bv}$ with 
$\gamma^{(1)}_{\bv}\in \Gamma_{\bv}(\gamma^{(0)})$ and $m_{\bv}\in M(\gamma^{(1)}_{\bv})$ 
taking into account non-unique parentage (i.e. there is no overcounting), 
that is, 
\be \label{2.6}  
\Psi=\sum_{{\bv}\in V}\;\sum_{l\in L_{\bv}}\; \kappa_{{\bv},l}\; \eta(T_l)
\ee

We define $C_v, \Psi_v, V_v, C_{v_1\wedge v_2}, \Psi_{v_1\wedge v_2}$ as follows:\\

Fix a vertex  $v \in V(\gamma^{(0)})$ and consider the set $C_v$ of elements $\eta (T_l)$ where $\eta(T_l) \in C_v$ iff there exists a
primordial parent based on $\gamma^{(0)}$ on which the action of the constraint at vertex $v$ produces, upto the action of a diffeomorphism, the state
$T_l$. 

We define the  restriction of $\Psi$ to a vertex $v \in V(\gamma^{(0)})$ to be the state
$\Psi_v$ obtained by setting to zero in (\ref{2.6}), the coefficients of all $\eta (T_l) \notin C_v$.

Define the set  $V_v\subset V(\gamma^{(0)})$ where $v^{\prime} \in V_v$ iff   $\Psi_v$ fails to annhilated by the constraint
action at $v^{\prime}$. 

Define the set $C_{v_1\wedge  v_2}$ to be the set of all $\eta (T_l)$ in (\ref{2.6})  for which $T_l$ (upto the action of a diffeomorphism) is produced by the action of the 
constraint at $v_1$ as well as at $v_2$. Define the restriction $\Psi_{v_1\wedge v_2}$  of $\Psi$ to be the state obtained by 
setting to zero in (\ref{2.6}), the coefficients of all $\eta (T_l) \notin  C_{v_1\wedge v_2}$.

\subsubsection*{Definition of Propagation}

If there exists $v\in V(\gamma^{(0)})$  such that $V_v \neq \emptyset$  we say that $\Psi$ encodes propagation. If $V_v =\emptyset$ for some $v\in V(\gamma^{(0)})$, 
we say that there is no propagation from $v$. If $V_v=\emptyset$ for every $v\in V(\gamma^{(0)})$, we say that the $\Psi$  does not encode propagation.

\subsubsection*{Definition of Propagation Distance}

Fix $v\in V(\gamma^{(0)})$ and consider any other $v^{\prime} \in V(\gamma^{(0)})$.  
If $\Psi_{v\wedge v^{\prime}}$  does not  solve the equations at $v^{\prime}$, we say that there is {\em  immediate propagation from $v$ to $v^{\prime}$} else that there
is no immediate propagation from $v$ to $v^{\prime}$. Note that there can be immediate propagation from $v$ to $v^{\prime}$ as well as from $v^{\prime}$ to $v$  and there could also be 
`one way' immediate propagation only from $v$ to $v^{\prime}$ but not vice versa.

Note also that 
$\Psi_{v\wedge v^{\prime}}$ solves the constraint equations
at $v^{\prime}$ for all $v^{\prime} \notin V_v$. This follows immediately 
from the fact  that (a) $\Psi_v$ solves these equations at such $v^{\prime}$ and (b) the equations for $\Psi_v$ at any $v^{\prime}\neq v$ only involve elements of $C_{v\wedge v^{\prime}}$.
It follows that for $v^{\prime} \notin V_v$ there is no immediate propagation from $v$ to $v^{\prime}$. It also follows from (b) and the definition of $V_v$ that there is immediate propagation
from $v$ to any element of $V_v$.

A chain ${\cal C}_{v_1,v_2}$ of propagation  of length $n$ from $v_1$ to $v_n$ is a set of vertices  $\{v_i\} \subset V(\gamma^{(0)})$ such that there immediate  propagation from $v_i$ to $v_{i+1}, i=1,..,n-1$.
If there exists  a chain ${\cal C}_{v_1,v_2}$ then we say that there is propagation from $v_1$ to $v_2$.  
If there is propagation from $v_1$ to $v_2$, the  {\em propagation distance} from $v_1$ to $v_2$ is the length of the shortest chain from $v_1$ to $v_2$.

\subsubsection*{Comments}
The various choices of $V$ (see the discussion before  (\ref{2.6}))
are a direct consequence of 
non-unique parentage
and thus any choice of $V$ as above is a valid choice capturing the idea 
that the corresponding term in (\ref{2.6}) is produced by the action of 
the constraint at $v$. Note however that the role of the set $V$ is only to 
provide an explicit
labelling index for the coefficients which appear
in  (\ref{2.6}). The set $V$ plays no role in the above 
definition of propagation.

The definition of propagation is based directly on the coupling  between the  constraint equations at different vertices. 
The intuitive picture  underlying the definition as formulated above is as follows.
We think  of a `perturbation' or `disturbance'  (roughly speaking the extraordinary edge between 2 vertices) as already being present in the solution $\Psi$ with  each child $\eta (T_l)$ thought of as the quantum analog of canonical
data on a slice. The way we determine how the perturbation  `evolves'  depends on how we view these `quantum slices'. 
Consider a vertex $v$ with the children obtained through the constraint action at $v$ being viewed as encoding a perturbation/disturbance/signal originating at $v$.
Suppose $v^{\prime} \in V_v$. Then this signal, encoded in the precise combination of children  with parental vertex  $v$, propagates through $v^{\prime}$ to a precise combination of children
with parental vertex $v^{\prime}$. If this signal was absent the effect of  its propagation (encoded in the precise coefficients of the children obtained through the insertion of extraordinary edges
on edges which do not connect $v^{\prime}$ to $v$) would also be absent (i.e. the combination of children  with parental vertex $v$  $\Psi_{v\wedge v^{\prime}}$ do not solve the equations 
at $v^{\prime}$ but need to be augmented by the combination of the remaining the children with parental vertex at $v^{\prime}$). Note that since $v^{\prime}$ shares some parentage with $v$, it must be 
connected to and hence an `immediate' neighbor of $v$. Hence the nomenclature `immediate propagation'. 
On the other hand if $v^{\prime} \notin V_v$, then either $v$ and $v^{\prime}$ are not nearest neighbors (so there is no question of immediate propagation) or they are nearest neighbors but the
presence of children with parental vertex $v$ does not prevent the precise combination of the rest of the children with parental vertex $v^{\prime}$ from solving the equations at $v^{\prime}$
i.e. not only does $\Psi_{v^{\prime}}$ automatically solve these equations but so does $\Psi_{v^{\prime}}- \Psi_{v\wedge v^{\prime}}$. 
In the latter case the disturbance between  $v$  and $v^{\prime}$ (encoded in $\Psi_{v\wedge v^{\prime}}$) may be viewed as not propagating beyond $v^{\prime}$.

Again, with more wording it is obvious how to generalise this to 
solutions of higher generation or to solutions which involve more than
one generation. In contrast to the presence of non-unique parentage
the presence of propagation does depend on the details of the dynamics as
we need access to explicit solutions to determine if they encode propagation.
These are difficult to construct  in a SU(2) calculation 
for the reason that the volume operator is not analytically diagonalisable.
Thus, one has to resort to numerical techniques. To motivate 
an in depth numerical analysis for SU(2),
we turn to the U(1)$^3$ theory in the next section and study an almost 
atomic example for which a multiparameter class of solutions can be constructed.

This example opens up the possibility of an alternate definition of propagation based 
on a class of solutions rather than a single fixed solution as above. 
As already mentioned, we shall discuss this 
alternate definition after we construct the class of solutions alluded to above.
Since the definition
in this section is based on a single solution we refer to it as a 
definition of {\em intrinsic}
propagation in contrast to the alternative, `space of solutions' 
dependent definition which we 
refer to as a definition of {\em extrinsic} propagation.  We emphasise 
again that propagation is 
a subtle notion and  its capture in a complete definition is expected to 
rely on experience with 
a variety of concrete examples. Consequently both the intrinsic definition 
and the extrinsic one
are to be taken as working definitions to be modified in response to inputs 
in the future.

\section{Demonstration of Existence, Non-Normalisabilty and Propagation 
for U(1)$^3$ Model}
\label{s1}

In this section we will study the question of existence, boundedness and 
propagation for a generation 1 class of solutions 
descending from a concrete and simple primordial $\gamma^{(0)}$.
We work within the 
$U(1)^3$ theory for which all constraint equations can be solved explicity.
Thus the colourings $m,n,..$ are now valued in $\mathbb{Z}^3$. Accordingly 
we speak of charge network functions (CNWF) rather than SNWF.

\subsection{The primordial parent graph}
\label{s1.1}

In what follows, we provide the necessary and sufficient information 
on the parent graph, such that the action of the Hamiltonian constraint 
is unambiguously defined:\\
\\
We consider a closed graph $\gamma_0$ with five vertices three of which are 
six-valent and two of which are four-valent. The graph is chosen so 
small that it fits into a single chart 
and, to be very explicit, we choose a right oriented coordinate frame
such that on a piece of paper the y axis points to the right,
the z axis upwards and the x axis towards the observer out of the sheet
of paper plane.  
The five 
vertices all lie on the y-axis and we label them $Z,A,B,C,D$ from left to 
right. There are two edges each between the pairs of vertices
$(Z,A)$, $(A,B)$, $(B,C)$, $(C,D)$ and $(D,Z)$ respectively and no others
that connect different vertices,
i.e. there are altogether ten such edges. We choose the pair of edges between 
$Z,A$ and $B,C$ respectively to lie in the y,z plane while we choose    
the pair of edges between 
$A,B$ and $C,D$ respectively to lie in the x,y plane. 
The pair of edges between $D,Z$ are such that the four edges at 
$Z$ and $D$ respectively have co-planar tangents there. 
All edges are 
semi-analytic and are oriented such that the respective pairs point from 
$Z$ to $A$, from $A$ to $B$, from $B$ to $C$, from $C$ to $D$ and from
$D$ to $Z$. In addition, at vertices $A,B,C$ there is a semi-analytic 
loop attached, intersecting the graph nowhere else as
specified below.

More specifically, 
the beginning segments of the pair of edges between
$D,Z$ lie in the x,y plane and their end segment in the y,z plane 
while in between they wind around accordingly without intersecting or 
knotting. The other edges also do not knot. The two edges between $Z,A$ and 
$B,C$ respectively are such that 
one of them  lies above (``up'') the x,y plane 
and the other below (``down''). Accordingly 
we label them $u_{ZA},d_{ZA}$ and $u_{BC},d_{BC}$ respectively. Likewise,
the two edges between $A,B$ and 
$C,D$ respectively are such that 
one of them lies above the y,z plane (``front'' of the sheet)  
and the other below (``back'' of the sheet). Accordingly 
we label them $f_{AB},b_{AB}$ and $f_{CD},b_{CD}$ respectively.
Finally, the pairs of edges $u_\ast,d_\ast,\; 
\ast\in \{ZA,BC\}$ intersect in both their beginning and final point at a 
non-vanishing angle (say $\pi/2$) and likewise 
the pairs of edges $f_\ast,b_\ast,\; 
\ast\in \{AB,CD\}$. The edges between $D,Z$ will be denoted $f_{DZ}, b_{DZ}$ 
respectively where $f_{DZ}$ is of f-type when leaving $D$ and of u-type when 
entering $Z$ while
$b_{DZ}$ is of b-type when leaving $D$ and of d-type when 
entering $Z$. The loops are labelled $l_A,\; l_B,\; l_C$ respectively and 
have the following properties: $l_A$ has a beginning analytic 
segment which is 
tangent to $u_{ZA}$ at $A$ and an end segment
tangent to $b_{AB}$ at $A$ and is semi-analytic 
in between. Likewise $l_B$ is tangent to $f_{AB}$ in $B$ in its beginning 
and to $d_{BC}$ at $B$ in its end while $l_C$ is tangent to $u_{BC}$ at $C$
in the beginning and to $b_{CD}$ at $C$ in its end. The loop $l_A$ is 
knotted while $l_B,l_C$ are unknotted.

\begin{figure}[hbt]
\includegraphics[width=18cm,height=12cm]{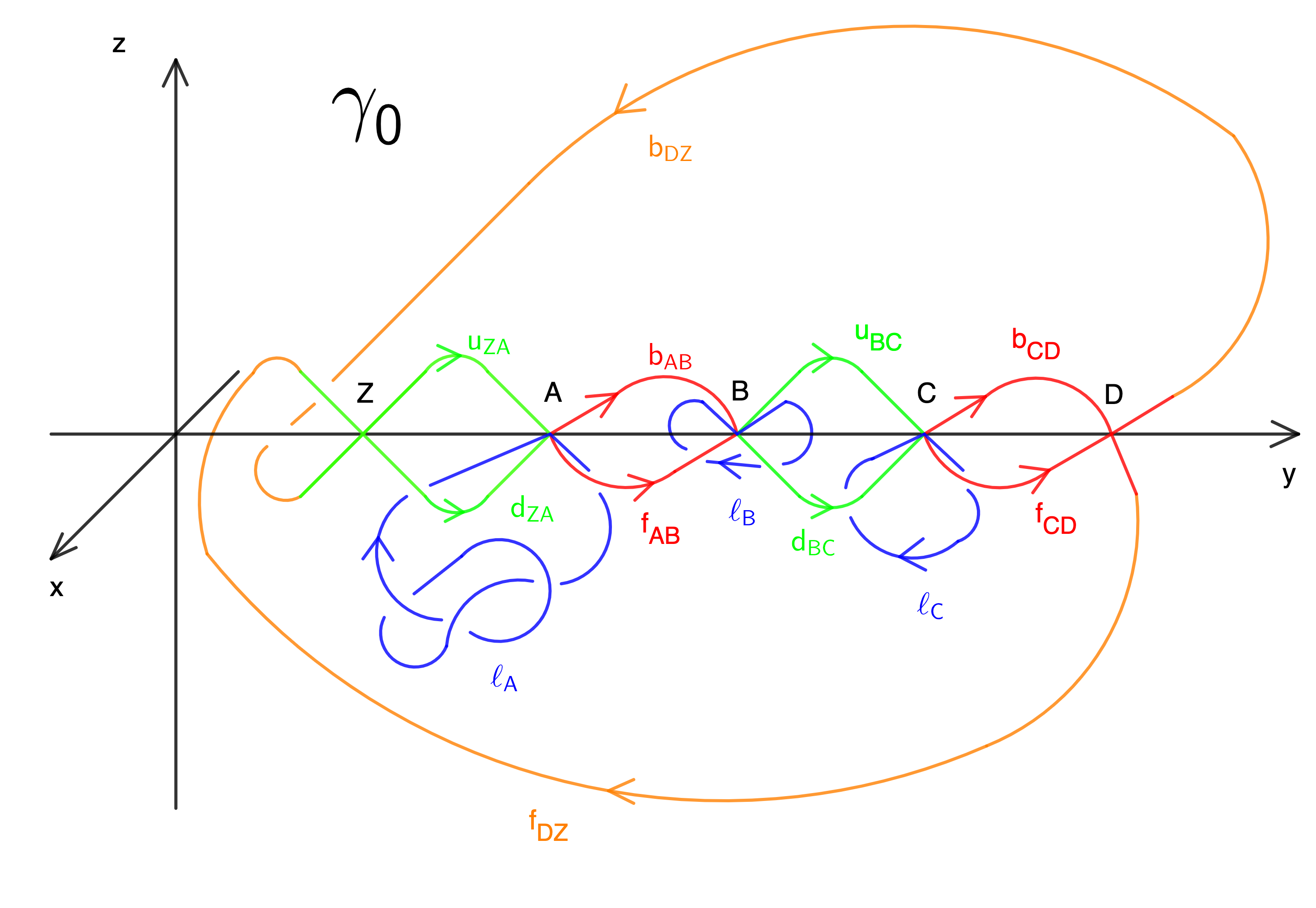}
\caption{The parent graph. Green lines are in the y,z plane and red lines in 
the x,y plane.
Orange lines interchange between these planes. The u (d) lines have positive
(negative) z coordinates while the f (b) lines have positive (negative) x 
coordinates. The blue loops are tangent to edges at their endpoints
as indicated.}
\end{figure}

\subsection{The parent charges}
\label{s1.2}

In the U(1)$^3$ theory, the edges of charge network functions (CNWF) carry 
a charge vector, i.e. a triple of integers. Note that the charge vector 
by definition is non-trivial, i.e. at least one of those integers is supposed
to be non-vanishing, otherwise the CNWF is defined over the smaller 
graph with the uncharged edge dropped. We denote the charge vectors
of the edges 
$u_{ZA},\; f_{AB}, \;u_{BC}, \; f_{CD}, \; b_{DZ}$ respectively by 
$m_{ZA},\; m_{AB}, \;m_{BC}, \; m_{CD}, \; m_{DZ}$ 
and of of the edges 
$d_{ZA},\; b_{AB}, \;d_{BC}, \; b_{CD}, \; f_{DZ}$ respectively by 
$n_{ZA},\; n_{AB}, \;n_{BC}, \; n_{CD}, \; n_{DZ}$. 
The charge vectors of the loops $l_A,\; l_B,\; l_C$ are denoted 
respectively by $c_A,\; c_B,\; c_C$. 

We only consider CNWF which are solutions to the U(1)$^3$ Gauss constraint
which imposes the constraint that for all $\ast\in\{ZA,AB,BC,CD,DZ\}$ 
the vector
\be \label{1.1}
N_\ast:=m_\ast+n_\ast\equiv N
\ee
is the same vector $N$ i.e. independent of the label $\ast$. There is no 
constraint on the loop charges.
Accordingly, the 
CNWF over $\gamma_0$ is unambigously labelled by nine vectors in 
$\mathbb{Z}^3$, say the five $m_\ast$, the three $c_\ast$ and $N$. 

\subsection{Definition of the U(1)$^3$ Hamiltonian constraint}
\label{s1.3}

The U(1)$^3$ analog of the (smeared)
Euclidean Hamiltonian constraint is defined 
on a general CNWF $T_{\gamma,m}$ up to a state vector independent factor by 
\ba \label{1.2}
C(f)\; T_{\gamma,m} &=& 
\sum_{v\in V(\gamma)}\; \frac{f(v)}{[2i]\;T(v)}\;
C_v\ T_{\gamma,m}
\nonumber\\
C_v &=& \sum_{e_1\cap e_2\cap e_3=v}\; C_{v,e_1,e_2,e_3}
\nonumber\\
C_{v,e_1,e_2,e_3} &=& \sum_{I,J,K,l\in \{1,2,3\}}\; \epsilon^{IJK}\; 
[h^l_{\alpha_{\gamma,v,e_I,e_J}}-
(h^l_{\alpha_{\gamma,v,e_I,e_J}})^{-1}]\;
h^l_{s_{\gamma,v,e_K}}\;
[V_v,(h^l_{s_{\gamma,v,e_K}})^{-1}]
\ea
where the notation is as follows: $V(\gamma)$ is the set of vertices 
of $\gamma$ and $E(\gamma)$ the set of its oriented edges. The natural 
number $T(v)$ is the number of unordered triples of edges adjacent to $v$ 
which have linearly independent tangents at $v$ and 
the second sum in (\ref{1.2}) is the sum over those triples of edges.
If an edge $e$ is adjacent to $v$ we define $s_{\gamma,v,e}$ to be 
a segment of $e$ connected to $v$ but not to the 
other endpoint of $e$ with an orientation outgoing from $v$. Given 
two edges $e,e'$ adjacent to $v$ we consider a loop 
$\alpha_{\gamma,v,e,e'}$ starting in $v$ along $s_{\gamma,v,e}$ and ending 
in $v$ along $s_{\gamma,v,e'}^{-1}$. To complete the loop, there 
is a connecting arc $a_{\gamma,v,e,e'}$ from the endpoint of 
$s_{\gamma,v,e}$ to the endpoint of $s_{\gamma,v,e'}$ which is unknotted 
and intersects $\gamma$ nowhere else. Thus 
\be \label{1.3}
\alpha_{\gamma,v,e,e'}=s_{\gamma,v,e}\circ a_{\gamma,v,e,e'}\circ 
s_{\gamma,v,e'}^{-1}
\ee
We require $a_{\gamma,v,e',e}=a_{\gamma,v,e,e'}^{-1}$ so that also
$\alpha_{\gamma,v,e',e}=\alpha_{\gamma,v,e,e'}^{-1}$. The ordering 
within the triple is such that the tangents of 
$s_{\gamma,v,e_1},s_{\gamma,v,e_2},s_{\gamma,v,e_3}$ at $v$ form
a right oriented basis if they are linearly independent. The routing 
of the arc through the edges incident at $v$ is described in detail
in \cite{QSDI} and reduces for the case that $e,e'$ lie in a 
coordinate plane to let the arc also lie in that plane.
Finally, 
by $h^j_p$ we denote the holonomy of the connection $A^j,\; j=1,2,3$ along 
the path $p$. 

The operator $V_v$ appearing in (\ref{1.2}) is the U(1)$^3$ analog 
of the Ashtekar-Lewandowski volume operator which reads explicitly
up to a state vector independent factor 
\be \label{1.4}
V_v=|\frac{1}{3!}\sum_{e_1\cap e_2 \cap e_3=v}\; \sigma(s_1,s_2,s_3) \; 
\epsilon_{jkl} \; X^j_{s_1}\; X^k_{s_2}\; X^l_{s_3}|^{1/2}
\ee
where $s_I$ is the shorthand for $s_{\gamma,v,e_I}$, the integer
$\sigma(s_1,s_2,s_3)$ is the sign of the determinant of the matrix of 
column vectors $(\dot{s}_1(0),\dot{s}_2(0),\dot{s}_3(0))$ and  
$X^j_s=-\;h^j_s\; \frac{\partial}{\partial h^j_s}$. Note that in (\ref{1.4}) 
we sum over all triples of edges, not only those whose tangents are 
linearly independent and not only those whose ordered tangents are right
oriented.  The whole purpose of considering 
the U(1)$^3$ truncation of the actual SU(2) theory is that (\ref{1.4}) 
is diagonal on CNWF while in the SU(2) theory $X^j_s$ is replaced by a 
right invariant vector field on SU(2) 
so that (\ref{1.4}) needs the spectral 
theorem for an explicit evaluation which except for specific spin 
configuration is not possible analytically.  
 
Specifically, suppose that an edge $e$ is adjacent to $v$, carries charge 
$m_e$ and is outgoing from (incoming to) $v$ so that 
$e=s^{z_{\gamma,v,e}}\circ e'$ with $z_{\gamma,v,e}=+1$ ($z_{\gamma,v,e}=-1$) 
where $e'$ is disjoint from $v$. Then $X^j_s$ 
is diagonal on the corresponding CNWF with eigenvalue 
$z_{\gamma,v,e}\; m_e^j$ 
and the eigenvalue of the volume operator is given explicitly by
\be \label{1.4a}
\nu_v=|Q_v|^{1/2},\;
Q_v=\frac{1}{3!}
\sum_{e_1\cap e_2\cap e_3=v} \; [\prod_{I=1}^3 z_I]\; \sigma(s_1,s_2,s_3)\;
\det(m_{e_1},m_{e_2},m_{e_3})
\ee   

\subsection{Notes on closability}
\label{s1.3a}

For completeness, although not necessary for the understanding of the 
rest of this paper, we note that the operator (\ref{1.2}) is not
symmetric in an obvious way. As mentioned, the arguments of \cite{Kuchar}
even suggest that the Hamiltonian constraint operator 
{\it must not be} symmetric. 
Nevertheless, the way it stands it is not even 
closable (i.e. its adjoint is densely defined) which would be 
a prerequisite for having a symmetric operator. This is because 
it may happen that in the decomposition into CNWF $T_{\gamma'}$ 
of its action on a CNWF $T_{\gamma}$ some $\gamma'$ appear that contain 
the extraordinary edges that the Hamiltonian constraint adds but not all of the 
beginning segments $s_{\gamma,v,e}$. Clearly,  this `edge disappearance' occurs 
only if  these segments in $T_{\gamma}$  happen to be labelled by charges which are
cancelled by the charges carried by the corresponding segments of the loops added by the 
constraint action.
Since such $T_{\gamma'}$ can be 
produced from an uncountably infinite number of mutually 
orthogonal $T_\gamma$ (e.g. all 
$\gamma$ that differ from a given $\gamma_0$ by deforming just 
$s_{\gamma_0,v,e}$ into $s_{\gamma,v,e}$) the adjoint of (\ref{1.2}) 
is not densely defined on CNWF. 

In the context of gauge group $SU(2)$ (i.e. full blown gravity rather than the $U(1)^3$ model)
this disappearance of edges with consequent non-closability can occur with the constraint action
constructed in the arXiv version of Reference \cite{QSDII} whenever the segments of $T_{\gamma}$
carry spin $\frac{1}{2}$. As indicated in Footnote  \ref{fn2} this phenomenon implies that the 
solutions tracing to a single unique primordial do not exhaust  
the space of solutions, contrary to
the assumption in that arXiv version (which 
was addressed in the published version). 
However, one may, as in the work in this paper on the $U(1)^3$ model, 
restrict attention to the class of solutions from a unique primordial. The results of this paper then
strongly indicate that even this restricted class contains propagating solutions. 

Various proposals have been made 
in the literature to make the operator closable in the $SU(2)$ case which apply immediately to the simpler context of $U(1)^3$.
Perhaps the minimal 
correction to (\ref{1.3}) that makes this possible is as follows (see
\cite{ST} for details): Simply substitute $H_{v,e_1,e_2,e_3}$ in (\ref{1.2})
by 
\be \label{1.4aa}
H'_{v,e_1,e_2,e_3}=
P_{v,e_1}\;P_{v,e_2}\;P_{v,e_3}\;H_{v,e_1,e_2,e_3},\;\;
P_{v,e_I}:=1-\theta(\Delta_{s_{\gamma,v,e_I}})
\ee
Here $\theta(x)=1$ for $x\ge 1$ and $\theta(x)=0$ otherwise is the step
function and $\Delta_{s}=\sum_j (X^j_s)^2$ the Laplacian for $s$. 
These spectral projections avoid the effect just described. The 
semiclassical properties of these projections are discussed in 
\cite{ST}, suffice it to say here that we may interpret $\Delta_s$ as the 
quantisation of $-\sum_j E_j(S_s)^2$ where $S_s$ is an arbitrarily 
small but finite surface intersection $s$ transversally. If the classical 
3-metric is non degenerate then $\theta(-\sum_j E_j(S)^2)=0$ and in that 
sense the modification is justified. 

With this modification, the operator becomes closable and could be 
symmetrically ordered, see the third reference of \cite{Books} for 
the technical statement. The conclusions of the rest 
of this paper apply to both the non closable version (\ref{1.2}) 
and the closable modification just discussed because they both exhibit 
the phenomenon of non-unique parentage. By contrast,
the closable Euclidean operators in the published version of \cite{QSDII} 
based on so-called
double kinks instead of the loops considered here 
or in \cite{LS} based on loops which intersect the original graph 
in only one vertex, both by design have unique parentage and therefore do 
not exhibit propagation.

Recently, the constraint action construction techniques of \cite{QSDI} have been 
combined with  key  geometrical  insights into the classical constraint action developed in \cite{aamv}.
The resulting constraint action \cite{MV}
\footnote{The first paper in Reference \cite{MV} constructs three such actions. The second paper seeks to demonstrate
anomaly free constraint commutators for one of them, referred to as the `Mixed Action' in the first. Our comments here  pertain to this
Mixed Action.}
differs from the ones mentioned hitherto in several ways. In particular the
loops added by the constraint are labelled by spin representations which are tailored to
the labels of the parent state being acted upon. As a result there are {\em generically} children
in which edge segments disappear. 
Since a detailed analysis of propagation for the constraint action of  \cite{MV}
constitutes an interesting but as yet open problem, we restrict ourselves to the following remarks:\\
(i) A preliminary analysis suggests that non-unique parentage is {\em primarily} associated with children
for which edges in the parent disappear.\\
(ii) As indicated in our  introductory remarks in section 1, while non-unique parentage invalidates the arguments of 
\cite{Smolin}, the existence of explicit solutions to the constraints involving children of non-unique parentage
must be established, a task which is still incomplete in the context of \cite{MV}.\\
(iii) In the context of `disappearing edges', non-unique parents are expected to be labelled, generically, by 
diffeomorphically distinct graphs. Since our definition of propagation in this work is based on a fixed primordial graph structure
it is necessary to generalise this definition appropriately. We note here that a qualitative and intuitive description of 
propagation in the context of variable parental graph structure is provided in \cite{MVprop} in the context of a novel $U(1)^3$ constraint
action distinct from the QSD type action of this paper.

\subsection{Evaluation of the Hamiltonian constraint on the chosen parent
CNWFs}
\label{s1.4}

By construction, the tangents of all edges adjacent to the vertices $Z,D$ 
are co-planar so that the operator $[V_v,(h_{s_{\gamma,v,e}})^{-1}]$ with 
$e$ adjacent to $v$ vanishes there. Accordingly, the Hamiltonian 
constraint has non-trivial action only at vertices $A,B,C$. At each 
of these vertices, the Hamiltonian constraint produces 
children graphs which correspond to gluing in an arc of the 
above type between pairs of adjacent edges. 
The non-uniqueness of parentage and the non-trivial correlation between 
the actions of the Hamiltonian constraint at different vertices when 
computing solutions to the Quantum Einstein Equations here can be 
me made explicit and transparent. Specifically, a child CNWF with 
an arc in between 
$A,B$ can come from either the action at $A$ or $B$ but originating
from two different CNWF over $\gamma_0$ that is, with different parental 
charges so that these two parental CNWF cannot be related by a diffeomorphism. 
The same applies to a child with an arc between $B,C$. Accordingly,
when constructing a solution to the Hamiltonian and spatial diffeomorphism 
constraint as a superposition of diffeomorphism averages of child CNWF of 
the above type, the coefficients in that superposition are constrained 
not by mutually disjoint sets of constraints, one set for each vertex,
but rather by a coupled system of equations. In what follows, we will make 
this explicit and show that the resulting system of equations admits 
non-trivial solutions.\\
\\
We will thus consider child graphs $\gamma_1$ and $\gamma_2$ respectively 
where $\gamma_1$ differs from the parental graph $\gamma_0$ by an arc $a_1$
between interior points of $f_{AB}$ and $b_{AB}$ in the x,y plane
respectively 
starting from the point on $f_{AB}$ while 
$\gamma_2$ differs from the parental graph $\gamma_0$ by an arc $a_2$
between interior points on $d_{BC}$ and $u_{BC}$ in the y,z plane 
starting from a point on $d_{BC}$. 
The action at $B$ on $\gamma_0$ produces both $\gamma_1$ and $\gamma_2$
types of graphs plus additional ones $\gamma'_B$
while the action at $A$ only produces the $\gamma_1$ 
type and additional ones $\gamma'_A$
and the action at $C$ only produces the $\gamma_2$ 
type 
and additional ones $\gamma'_C$.
In what follows the $\gamma_A', \gamma_B',\gamma'_C$ contributions 
to the Hamiltonian constraint will be ignored because we construct 
distributional solutions 
out of diffeomorphism averages of CNWF over $\gamma_1,\gamma_2$ which, as 
distributions, have 
trivial action on CNWF over $\gamma'_\ast,\;\ast\in\{A,B,C\}$.\\
\\
{\bf Action at $A$:}\\
We notice that with the orientation choices 
made, the following triples of edges adjacent at and outgoing 
from $A$ are such that their 
tangents form a right oriented triple of linearly independent vectors at $A$
\be \label{1.5}
(f_{AB}, b_{AB}, u_{ZA}^{-1}),\;
(b_{AB}, f_{AB}, d_{ZA}^{-1}),\;
(d_{ZA}^{-1},u_{ZA}^{-1},b_{AB}),\;
(u_{ZA}^{-1},d_{ZA}^{-1},f_{AB})
\ee
The orientations of those edges coincides with those of the outgoing segments
from $A$ whence 
$z_{\gamma_0,A,u_{ZA}}=z_{\gamma_0,A,d_{ZA}}=-1$ and 
$z_{\gamma_0,A,f_{AB}}=z_{\gamma,A,b_{AB}}=+1$. Thus the volume eigenvalue
at $A$ derives as $\nu_A=|Q_A|^{1/2}$ with
\ba \label{1.6}
Q_A &=& Q(m_{ZA},n_{ZA},m_{AB},n_{AB},c_A,c'_A)_{c'_A=c_A};\;\;
Q(m_{ZA},n_{ZA},m_{AB},n_{AB},c_A,c'_A)
\\
&=&
-[\det(m_{AB},n_{AB}+c^{\prime}_A,m_{ZA}+c_A)+\det(n_{AB}+c^{\prime}_A,m_{AB},n_{ZA})]  \nonumber\\ 
&&+[\det(n_{ZA},m_{ZA}+c_A,n_{AB}+c'_A)+\det(m_{ZA}+c_A,n_{ZA},m_{AB})]   
\nonumber
\ea
where the factor $3!$ got cancelled because permutation of edges within 
the above triples all give the same contribution. We have exploited 
that while the tangents of $u_{ZA}, l_A$ point into the same direction
at $A$ for the beginning segment of $l_A$, the loop is here outgoing while 
$u_{ZA}$ is ingoing. Likewise, while the tangents of 
$b_{AB}, l_A$ point into the same directions
at $A$ for the end segment of $l_A$, the loop is here ingoing while 
$b_{AB}$ is outgoing. 
The number of arguments of the function $Q$ 
is redundant due to (\ref{1.1}) but we will keep it for 
reasons of more transparent bookkeeping. We will use $\nu=\sqrt{|Q|}$ 
in what follows.

Thus the $\gamma_1$ type contribution of the Hamiltonian constraint $C_A$ at 
vertex $A$ is given by (we drop the common factor $T(v)=12,\; v=A,B,C$ 
in what follows as it can be absorbed into the lapse function and 
$m$ stands collectively for all charges on all edges of $\gamma_0$)
\ba \label{1.7}
&& C_A\; T_{\gamma_0,m}=\sum_l\;
\{
[\nu(m_{ZA}+\delta_l,n_{ZA},m_{AB},n_{AB},c_A,c_A)
-\nu(m_{ZA},n_{ZA},m_{AB},n_{AB},c_A,c_A)]\;
[h^l_{\alpha_A}-(h^l_{\alpha_A})^{-1}]\; 
\nonumber\\
&& +[\nu(m_{ZA},n_{ZA}+\delta_l,m_{AB},n_{AB},c_A,c_A)
+\nu(m_{ZA},n_{ZA},m_{AB},n_{AB},c_A-\delta_l,c_A)
\nonumber\\
&& -2\nu(m_{ZA},n_{ZA},m_{AB},n_{AB},c_A,c_A)]\;
[(h^l_{\alpha_A})^{-1}-h^l_{\alpha_A}]\;\} 
T_{\gamma_0,m}
\nonumber\\
&=& \sum_l\;
[\nu(m_{ZA}+\delta_l,n_{ZA},m_{AB},n_{AB},c_A,c_A)
-\nu(m_{ZA},n_{ZA},m_{AB},n_{AB},c_A-\delta_l,c_A)
\nonumber\\
&& -\nu(m_{ZA},n_{ZA}+\delta_l,m_{AB},n_{AB},c_A,c_A)
+\nu(m_{ZA},n_{ZA},m_{AB},n_{AB},c_A,c_A)]\;
[(h^l_{\alpha_A})^{-1}-h^l_{\alpha_A}]\;\} 
T_{\gamma_0,m}
\nonumber\\
&=:& \sum_l\; d^l_A(m_{ZA},n_{ZA},m_{AB},n_{AB},c_A)\;
[h^l_{\alpha_A}-(h^l_{\alpha_A})^{-1}]\; 
T_{\gamma_0,m}
\ea
where $\alpha_A=s_{\gamma_0,A, f_{AB}}\circ a_1\circ 
s_{\gamma_0,A, b_{AB}}^{-1}$ and $\delta_l\in \mathbb{Z}^3$ is the vector 
with component $[\delta_l]^j=\delta_l^j$. In (\ref{1.7}) we have 
considered the beginning segements $s$ of the 
edges $u_{ZA}, d_{ZA},l_A$ 
that have linearly independent tangents at $A$ together 
with the tangents of $f_{AB}, b_{AB}$ (that is why there is no contribution
from the end segment of $l_A$). Their holonomies along $s$ 
with outgoing orientation and charge $\pm\delta_l$ enters the commutator 
with the 
volume operator explaining the argument shifts by $=\pm\delta_l$ 
in the functions $\nu$ displayed. Then the loop $\alpha_A$ or $\alpha_A^{-1}$
respectively gets attached when the tangents of $s,f_{AB}, b_{AB}$ in this 
order are right or left oriented respectively.  
\\
\\   
{\bf Action at $C$:}\\
The situation here with respect to orientations of edges 
and charges is exactly as at $A$ with the substitutions
of labellings $ZA\to BC$ and $AB\to CD$ and $A\to C$. Therefore
\be \label{1.8}
Q_C=Q(m_{BC},n_{BC},m_{CD},n_{CD},c_C,c_C')_{c_C'=c_C}
\ee

Thus the $\gamma_2$ type contribution of the Hamiltonian constraint $C_C$ at 
vertex $C$ is given by 
\ba \label{1.9}
&&C_C\; T_{\gamma_0,m}=\sum_l\;
\{
[\nu(m_{BC},n_{BC},m_{CD},n_{CD}-\delta_l,c_C,c_C)
-\nu(m_{BC},n_{BC},m_{CD},n_{CD},c_C,c_C)]\;
[h^l_{\alpha_C}-(h^l_{\alpha_C})^{-1}]
\nonumber\\
&& +[\nu(m_{BC},n_{BC},m_{CD}-\delta_l,n_{CD},c_C,c_C)
+\nu(m_{BC},n_{BC},m_{CD},n_{CD},c_C,c_C+\delta_l)
\nonumber\\
&& -2\nu(m_{BC},n_{BC},m_{CD},n_{CD},c_C,c_C)]\;
[(h^l_{\alpha_C})^{-1}-(h^l_{\alpha_C})]\}\;
T_{\gamma_0,m}
\nonumber\\
&=& \sum_l\;
[\nu(m_{BC},n_{BC},m_{CD},n_{CD}-\delta_l,c_C,c_C)
-\nu(m_{BC},n_{BC},m_{CD},n_{CD},c_C,c_C+\delta_l)
\nonumber\\
&&
-\nu(m_{BC},n_{BC},m_{CD}-\delta_l,n_{CD},c_C,c_C)
+\nu(m_{BC},n_{BC},m_{CD},n_{CD},c_C,c_C)]\;
[h^l_{\alpha_C}-(h^l_{\alpha_C})^{-1}]\;T_{\gamma_0,m}
\nonumber\\
&=:& \sum_l\; d^l_C(m_{BC},n_{BC},m_{CD},n_{CD},c_C)\;
[h^l_{\alpha_C}-(h^l_{\alpha_C})^{-1}]\; 
T_{\gamma_0,m}
\ea
where $\alpha_C=s_{\gamma_0,C, d_{BC}}\circ a_2\circ 
s_{\gamma_0,C, u_{BC}}^{-1}$. We have exploited that it is now the 
end segment of $l_C$ which contributes.\\
\\
{\bf Action at $B$:}\\
The tangents of the 
following triples of edges are right oriented and outgoing from $B$
\be \label{1.10}
(d_{BC},u_{BC},f_{AB}^{-1}),\;
(u_{BC},d_{BC},b_{AB}^{-1}),\;
(f_{AB}^{-1},b_{AB}^{-1},d_{BC}),\;
(b_{AB}^{-1},f_{AB}^{-1},u_{BC}),\;
\ee
The orientations of those edges coincides with those of the outgoing segments
from $B$ whence 
$z_{\gamma_0,B,f_{AB}}=z_{\gamma_0,B,b_{AB}}=-1$ and 
$z_{\gamma_0,B,u_{BC}}=z_{\gamma_0,B,d_{BC}}=+1$. Thus the volume eigenvalue
at $B$ derives as $\nu=|Q|^{1/2}$ where 
\be \label{1.11}
Q_B=Q(m_{AB},n_{AB},m_{BC},n_{BC},c_B,c_B')_{c_B'=c_B}
\ee
where it was exploited that while the tangents of $f_{AB}$ and the 
beginning segment of $l_B$ as well as the tangents of $d_{BC}$ and the 
end segment of $l_B$ point into the same direction at $B$, 
$f_{AB}$ is ingoing while the beginng segment of $l_B$ is outgoing from
$B$ and $d_{BC}$ is outgoing while the end segment of $l_B$ is ingoing
at $B$. 

Thus, following the same arguments as at vertices $A,C$  
the $\gamma_1$ type contribution of the Hamiltonian constraint 
at $B$ is given by 
\ba \label{1.12}
&& C^1_B\; T_{\gamma_0,m} 
= \sum_l\;
\{
[\nu(m_{AB},n_{AB},m_{BC},n_{BC}-\delta_l,c_B,c_B)
-\nu(m_{AB},n_{AB},m_{BC},n_{BC})]\;
[h^l_{\alpha^1_B}-(h^l_{\alpha^1_B})^{-1}]\;
\nonumber\\
&& +[\nu(m_{AB},n_{AB},m_{BC}-\delta_l,n_{BC},c_B,c_B)
+\nu(m_{AB},n_{AB},m_{BC},n_{BC},c_B,c_B+\delta_l)
\nonumber\\
&& -2\nu(m_{AB},n_{AB},m_{BC},n_{BC},c_B,c_B)]\;
[(h^l_{\alpha^1_B})^{-1}-(h^l_{\alpha^1_B})]\}\;
T_{\gamma_0,m}
\nonumber\\
&=& \sum_l\;
[\nu(m_{AB},n_{AB},m_{BC},n_{BC}-\delta_l,c_B,c_B)
-\nu(m_{AB},n_{AB},m_{BC},n_{BC},c_B,c_B+\delta_l)
\nonumber\\
&&
-\nu(m_{AB},n_{AB},m_{BC}-\delta_l,n_{BC},c_B,c_B)
_+\nu(m_{AB},n_{AB},m_{BC},n_{BC},c_B,c_B)]\;
[h^l_{\alpha^1_B}-(h^l_{\alpha^1_B})^{-1}]\;
T_{\gamma_0,m}
\nonumber\\
&=:& \sum_l \; d_{B,1}^l(m_{AB},n_{AB},m_{BC},n_{BC},c_B)\;
[h^l_{\alpha^1_B}-(h^l_{\alpha^1_B})^{-1}]\;
T_{\gamma_0,m}
\ea
where 
$\alpha^1_B=s_{\gamma_0,B,f_{AB}}^{-1}\circ a_1\circ s_{\gamma_0,v,b_{AB}}$.
We are abusing the notation since 
at the level of CNWF the arc $a_1$ coming from $B$ is generally
different from the one coming from $A$ but after diffeomorphism 
averaging they get identified and this is what matters in what follows.

Likewise the $\gamma_2$ type contribution of the Hamiltonian constraint 
at $B$ is given by 
\ba \label{1.13}
&& C^2_B\; T_{\gamma_0,m} 
= \sum_l\;
\{
[\nu(m_{AB}+\delta_l,n_{AB},m_{BC},n_{BC},c_B,c_B)
-\nu(m_{AB},n_{AB},m_{BC},n_{BC},c_B,c_B)]\;
[h^l_{\alpha^2_B}-(h^l_{\alpha^2_B})^{-1}]\;
\nonumber\\
&& +[\nu(m_{AB},n_{AB}+\delta_l,m_{BC},n_{BC})
+\nu(m_{AB},n_{AB},m_{BC},n_{BC},c_B-\delta_l,c_B)]\;
\nonumber\\
&&
-2\nu(m_{AB},n_{AB},m_{BC},n_{BC},c_B,c_B)]\;
[(h^l_{\alpha^2_B})^{-1}-h^l_{\alpha^2_B}]\}\;
T_{\gamma_0,m}
\nonumber\\
&=& \sum_l\;
[\nu(m_{AB}+\delta_l,n_{AB},m_{BC},n_{BC},c_B,c_B)
-\nu(m_{AB},n_{AB}+\delta_l,m_{BC},n_{BC},c_B,c_B)
\nonumber\\
&&
-\nu(m_{AB},n_{AB},m_{BC},n_{BC},c_B-\delta_l,c_B)
+\nu(m_{AB},n_{AB},m_{BC},n_{BC},c_B,c_B)
]\;
[h^l_{\alpha^2_B}-(h^l_{\alpha^2_B})^{-1}]\;
T_{\gamma_0,m}
\nonumber\\
&=:& \sum_l\;d_{B,2}^l(m_{AB},n_{AB},m_{BC},n_{BC})\;
[h^l_{\alpha^2_B}-(h^l_{\alpha^2_B})^{-1}]\;
T_{\gamma_0,m}
\ea
where 
$\alpha^2_B=s_{\gamma_0,B,d_{BC}}\circ a_2\circ s_{\gamma_0,v,u_{BC}}^{-1}$.
Again, at the level of CNWF the arc $a_2$ coming from $B$ is generally
different from the one coming from $C$ but after diffeomorphism 
averaging they get identified.

\subsection{Properties of the coeffcients $d^l_\ast$}
\label{s1.4a}

The twelve coefficients $d^l_A,\; d^l_C,\; d^l_{B,I};\; I=1,2$ computed
in the previous subsection are quite complicated. They have the following
general structure: They are linear combinations with coefficients
$\pm 1$ of the differences
\be \label{1.13a}
\nu_\ast(m+\sigma \delta_l,n,c)-\nu_\ast(m,n,c),\;
\nu_\ast(m,n+\sigma \delta_l,c)-\nu_\ast(\sigma \delta_l,n,c),\;
\nu_\ast(m,n,c+\sigma \delta_l)-\nu_\ast(m,n,c)
\ee
where $\ast\in \{A,C,(B,1),(B,2)\}$ and $\sigma=\pm 1$. For each value
of $\ast$ the function $\nu_\ast$ is a function of five charges as follows:
Respectively, the variables $m,n,c$ coincide with the charges
$m=m_{ZA},\;n=n_{ZA},\; c=c_A$ for $\ast=A$ with $m_{AB}, n_{AB}$ fixed,
$m=m_{CD},\;n=n_{CD},\; c=c_C$ for $\ast=C$ with $m_{BC}, n_{BC}$ fixed,
$m=m_{BC},\;n=n_{BC},\; c=c_B$ for $\ast=(B,1)$ with $m_{AB}, n_{AB}$ fixed,
$m=m_{AB},\;n=n_{AB},\; c=c_B$ for $\ast=(B,2)$ with $m_{BC}, n_{BC}$ fixed.
The functions $\nu_\ast$ themselves are square roots of the modulus of
a linear combinations with coefficients $\pm 1$ of four determinants
built from the five charge vectors involved or equivalently it is the
fourth root of a positive bilinear expression in those four determinants
with coeffcients $1, \pm 2$. The more explict form of those coefficients
can be found in appendix \ref{seca1}.

In what follows, the charges
$\Delta:=\{m_{ZA}, m_{CD}, m_{ZD}, c_A, c_B, c_C, N\}$
will be fixed and $n_\ast=N-m_\ast$ with $\ast=ZA, AB, BC, CD, DZ$ due to
gauge invariance so that the only variables are $m_{AB}, m_{BC}$.
We consider the vector ``fields'' $\mathbb{Z}^3\to \mathbb{R}^3$ given by
$m_{AB}\mapsto d_A(m_{AB}),\;
m_{BC}\mapsto d_C(m_{BC})$ and $\mathbb{Z}^6\to \mathbb{R}^3$ given by
$(m_{AB},m_{BC})\mapsto d_{B,I}(m_{AB},m_{BC});\;I=1,2$.
What we want to show is that all vector fields $d_A,d_C,d_{B,I}$ are
non-vanishing at ``generic points'' and moreover that $d_{B,1},d_{B,2}$
are linearly independent at ``generic points''. By generic points we mean
all of $\mathbb{Z}^3$ or $\mathbb{Z}^6$ respectively except for ``lower
dimensional'' sublattices of $\mathbb{Z}^3$ or $\mathbb{Z}^6$ respectively,
or even better on finite subsets of those respectively.
Here a lattice of dimension $k\le l$ in $\mathbb{Z}^l$ is a subset of the
form $a_1 e_1+..+a_k e_k$ with variable $a_1,..,a_k\in \mathbb{Z}$
and fixed linearly independent (over $\mathbb{Z}$)
$e_1,..,e_k\in \mathbb{Z}^l$. This would imply that 1. $d_A, d_C$ are
non-vanishing ``almost everywhere'' and thus are not compactly supported
and 2. that $d_{B,1}, d_{B,2}$ are linearly independent ``almost everywhere''.

We will not be able to give a strict proof of these statements but we will
give arguments for why this is very plausible. The subsequent discussion
will also show why it is difficult to turn the plausibility arguments
into a strict proof: We enter difficult questions of algebraic
geometry which can be addressed in principle but require much more work.
The arising questions and conjectured answers could benefit from machine
learning techniques which we advertise at this point.

To begin with, the
vector condition $d_A=0$ is a set of three equations for three unknowns
$m_{AB},\; l=1,2,3$ and thus will typically have only a finite number of
solutions even if we allow $m_{AB}\in \mathbb{C}^3$. The catch is in the word
``typically''. In principle it could happen that $d_A$ depends on $m_{AB}$ in
a sufficiently degenerate way so that the statements that we would like
to prove do not hold although this is intuitively hard to imagine.
The reason why this appears unlikely is as follows:
To investigate these questions analytically one would for instance (extending
$m_{AB}$ to $\mathbb{R}^3$) compute the Hessian
$H_A=\det(\frac{\partial d_A}{\partial m_{AB}})$ from whose zeroes one
would infer where the map $d_A$ fails to be injective. However, computing
that Hessian and its zeroes is even more difficult than determining the
zeroes  of $d_A$ directly. In order to say more than ``by inspection
$d_A$ depends on $m_{AB}$ sufficiently non-degenerately'', we consider solving
the equations $d_A^l=0$. We may write
(see appendix) $d_A^l=\nu^l_1+\nu^l_2-\nu^l_3-\nu^l_4$ where
$\nu^l_\mu=|Q^l_\mu|^{1/2},\;\mu=1,2,3,4$
and $Q^l_\mu$ has the form $a^{l,\mu}_j m_{AB}^j+b^{l,\mu}$
where $a^\mu_j, b^\mu$ are integers depending on $\Delta$. By inspection
(see appendix \ref{seca1}) the co-normals $\vec{a}^\mu\in \mathbb{Z}^3$ are
different for different $\mu$ because they involve charge shifts in different
entries of the determinants involved, hence the affine polynomials
$Q^l_\mu$ have different level hypersurfaces. Accordingly, for each $l$
these four polynomials $Q^l_\mu$ and thus the $\nu^l_\mu$ are algebraically
independent and the fact that the level hypersurfaces are different
suggests that the four $\nu^l_\mu$ cancel for each $l$ at most at
finitely many points.

To investigate this further, dropping the index
$l$ for notational simplicity, we obtain the condition
$\nu_1+\nu_2=\nu_3+\nu_4$. As all $\nu$ are not negative this is equivalent
with
\be \label{1.13b}
\nu_3^2+\nu_4^2-\nu_1^2-\nu_2^2=2(\nu_1 \nu_2-\nu_3\nu_4)
\ee
which implies upon squaring
\be \label{1.13c}
-(\nu_3^2+\nu_4^2-\nu_1^2-\nu_2^2)^2+4\;(\nu_1^2 \nu_2^2+\nu_3^2\nu_4^2)
=8\nu_1 \nu_2\nu_3\nu_4
\ee
and squaring again once more thereby introducing $q_\mu:=|Q_\mu|=\nu_\mu^2$
\be \label{1.13d}
[4 (q_1 q_2+q_3 q_4)-(q_1+q_2-q_3-q_4)^2]^2=64\; q_1 \; q_2\; q_3\; q_4
\ee
This equation is still not a polynomial in $m_{AB}$ because of the modulus
involved and one cannot get rid of it by squaring once more, the equation
is simply {\it not of algebraic type}. In order to turn it into a
polynomial, we
write $q_\mu=\epsilon_\mu Q_\mu$ with $\epsilon_\mu\in \{0,\pm 1\}$. Then,
at fixed $\epsilon_\mu$ (\ref{1.13d}) becomes a quartic polynomial in
$m_{AB}$. Suppose one finds a solution of (\ref{1.13d}) at fixed
$\epsilon_\mu$. Then, to make it a valid solution one has to check
that $\epsilon_\mu Q_\mu(m_{AB})>0$ if $\epsilon_\mu\not=0$ and
$Q_\mu(m_{AB})=0$ for $\epsilon_\mu=0$.

With this understanding, we are left with solving the following problem:
Let $x=m_{AB}^1,\;y=m_{AB}^2,\;z= m_{AB}^3$. Then for each choice of
$\epsilon_\mu$ ($3^4=81$ possibilities) we are looking for the zeroes
$(x,y,z)\in \mathbb{Z}^3$ of a non-homogeneous quartic polynomial
$P(x,y,z)$ with integer coefficients. In the theory of algebraic
geometry, this known as a {\it diophantine equation}. The condition
$P(x,y,z)=0$ defines an algebraic variety (surface) in $\mathbb{R}^3$ and
we are looking for its integral points. We simplify the problem and look
at the quartic polynomials $p_0(x,y):=P(x,y,0)$ with coefficients
in $\mathbb{Z}$ and for
$z\not=0$ at $p_z(X,Y):=P(x,y,z)/z^4,\; X=x/z,\;Y=y/z$ with coefficients
in $\mathbb{Q}$
which defines agebraic curves $C$. For this situation we
have the following deep statement in number theory.
\begin{Theorem}[Falting] ~\\
Let C be a non-singular algebraic curve over the rationals
(i.e. a curve in the plane defined by
the zeroes of a polynomial equation in x,y with rational coefficients
and such that
it is not self intersecting). If the curve has algebraic genus $g>1$ then
$\mathbb{Q}^2 \cap C$ is finite.
Here the algebraic genus for non singular
$C$ is g=(n-1)(n-2)/2 where $n$ is the degree of the polynomial.
\end{Theorem}
The theorem takes a step towards proving that
Fermat's famous equation $x^n+y^n=z^n$ has at most
finitely many positive integer solutions $x,y,z>0$ when $n\ge 3$: After
dividing by $z$ this becomes $X^n+Y^n=1$ which has at most finitely many
rational solutions $X=p/q,\; Y=r/s$ with $p,q$ and $r,s$ relative prime.
Then the solution to the original problem is $x=a(ps), y=a(rq), z=a(qs),\;a\in
\mathbb{N}$. Note that any homogeneous equation like Fermat's
has infinitely many if it has one, which is not true for
non-homogeneos equations,
and the hard part of the proof is thus to show that there is not a single one.
The algebraic
genus can be extended to algebraic curves with singularities and is
generically reduced by them \cite{Diophantine}.

Note that instead of introducing the $\epsilon_\mu$ one could momentarily
drop the requirement that the integers $q_\mu$ are not negative and
consider the solutions of the quartic (\ref{1.13d}). If one could argue
as above and conclude that the quartic has only finitely many solutions,
one would then conclude that it also has only a finite number of solutions
with $q_\mu\ge 0$. Then for each of these left over configurations one
would be left with solving the twelve equations
$Q_\mu^l=\pm q_\mu^l$ where we reintroduced the index $l$. As the $Q_\mu^l$
are algebraically independent, this system is again very unlikely to have
more than a finite number of solutions. In order to run this argument,
one would need to have an extension of Falting's
theorem at one's disposal for algebraic surfaces rather than curves
because (\ref{1.13d}) involves four $q_\mu$ rather than three $m_{AB}^l$.
(and for what follows more generally higher dimensional algebraic varieties
defined by integer coefficient polynomials). This would also ease the
proof in terms of $m_{AB}$ so that one does not have to
argue via the curves. Unfortunately, we were not able to find
such an extension in the literature.

Applied to our problem, assuming that the
algebraic curves in question are singularity free, we would conclude that
we obtain only a finite number of solutions of $p_0(x,y)=0$ for $z=0$ and
of $p_z(X,Y)=0$ for $z\not=0$
and thus most only finitely many 1-dimensional sublattices of $\mathbb{Z}^3$.
as solutions (note that only those of these solutions are admissable
for which the
above condition involving the $\epsilon_\mu$ is met).
However note that this is true for each direction $l$ in charge
space separately. As these finitely many discrete lines for different $l$
are unlikely to coincide, we expect that there are at
most finitely many solutions
of $d_A=0$. The same reasoning applies to $d_C$.

Another qualitative argument would be as follows: Compute the polynomials
$P_l(x,y,z)=0$ as above for each $l$ and each choice of the
$\epsilon_\mu$. These define 3 distinct algebraic surfaces.
The intesection of two
surfaces is one dimensional unless the two polynomials coincide which they
do not (if they would intersect in a common face then by analyticity they
would coincide everywhere). The intersection with the third is then also
at most one dimensional, more likely a discrete set of
points. From that curve or set one would need to pick the integer
points if they exist at all.

To apply this number theoretic argument to $d_{B,1}, d_{B,2}$ is significantly
harder. As for the vanishing of $d_{B,I}$ this again leads to a quartic
equation but now for a five dimensional (generalised - as it is no longer
defined by an algebraic equation) variety . Using the above heuristic
one would argue that each $d_{B,I}$ vanishes at most on the integer points of
a four dimensional, more likely three dimensional, variety $S_I$
which is different
for the two choices of $I$. For the points outside of the union $S=S_1\cup S_2$
of these sets the condition that
$d_{B,1}, d_{B,2}$ be co-linear can be stated as
$d_{B,1}\times d_{B,2}=0$. On $\mathbb{Z}^6-S$ only
two of these three equations are algebraically independent.
However, this time it is not possible to cast these into two algebraic
equations in six variables by squaring the equations sufficiently often.
The equation to be solved is simply not algebraic any more and almost
nothing can be said about its solution structure.

Yet, one expects that these two conditions single out an at most
four dimensional generalised 
variety of which one would again need to determine the integer points which
then very likely reduces this to a finite set.\\
\\
In summary, the analysis of
the singularity structure of the vector fields $d_A,d_C,d_{B,I}$
leads to hard questions in number theory and algebraic geometry which are
beyond the scope of this paper. However, the heuristic arguments given,
backed
up by some results from algebraic geometry make it appear extremely likely
that $d_A, d_B$ are nowehere vanishing and that $d_{B,1}, d_{B,2}$ are
nowhere linearly dependent, except on lower dimensional sublattices and
more likely on finite subsets. In other words, the violation of these
conditions, if possible at all, appears to be a tremendous number theoretic
accident and moreover one can exploit the freedom in the choice of $\Delta$
to downsize this set of violating points even further.

We will use the
assumption
of non singularity and linear independence of the vector fields at
generic points in the sense
described as a plausible conjecture in what follows.
To attempt a strict proof will require deeper methods from algebraic geometry
and number theory and could benefit from numerical machine learning
techniques.

\subsection{Decomposition into CNWF}
\label{s1.5}

We must
write the functions $[h^l_\beta-(h^l_\beta)^{-1}]\; T_{\gamma_0,m}$ 
in terms of CNWF 
for the loops $\beta=\alpha_A,\alpha_B^I,\alpha_C,\;I=1,2$ that appear 
in (\ref{1.7}), (\ref{1.9}), (\ref{1.12}) and (\ref{1.13}). 
For $\gamma_1$ 
we see that the edges $f_{AB},\;b_{AB}$ are split in halves 
$f_{AB}=\tilde{f}_{AB}\circ \hat{f}_{AB},\;
b_{AB}=\tilde{b}_{AB}\circ \hat{b}_{AB}$ and that the arc $a_1$ runs from 
$\tilde{f}_{AB}\cap\hat{f}_{AB}$ to 
$\tilde{b}_{AB}\cap\hat{b}_{AB}$. We thus need to introduce 
charge labels $\tilde{m}_{AB},\hat{m}_{AB},\tilde{n}_{AB},\hat{n}_{AB},m_1$
for $\tilde{f}_{AB},\;\hat{f}_{AB},,\;
\tilde{b}_{AB},\;\hat{b}_{AB},\;a_1$. Likewise,
for $\gamma_2$ 
we see that the edges $u_{BC},\;d_{BC}$ are split in halves 
$u_{BC}=\tilde{u}_{BC}\circ \hat{u}_{BC},\;
d_{BC}=\tilde{d}_{BC}\circ \hat{d}_{BC}$ and that the arc $a_2$ runs from 
$\tilde{d}_{BC}\cap\hat{d}_{BC}$ to 
$\tilde{u}_{BC}\cap\hat{u}_{BC}$. We thus need to introduce 
charge labels $\tilde{m}_{BC},\hat{m}_{BC},\tilde{n}_{BC},\hat{n}_{BC},m_2$
for $\tilde{u}_{BC},\;\hat{u}_{BC},\;
\tilde{d}_{BC},\;\hat{d}_{BC},\;a_2$ respectively.
Due to gauge invariance, these charges are subject to the constraints
\be \label{1.14}
\tilde{m}_{AB}=m_1+\hat{m}_{AB},\;
\tilde{n}_{AB}=-m_1+\hat{n}_{AB},\;
\tilde{m}_{BC}=-m_2+\hat{m}_{BC},\;
\tilde{n}_{BC}=m_2+\hat{n}_{BC},\;
\ee
The factor 
$(h^l_{\alpha_A})^\sigma,\; \sigma=\pm 1$ changes the charges 
$\tilde{m}_{AB}=\hat{m}_{AB}=m_{AB},\;
\tilde{n}_{AB}=\hat{n}_{AB}=n_{AB},\;m_1=0$
of $T_{\gamma_0,m}$ to 
\be \label{1.15}
\tilde{m}_{AB}=m_{AB}+\sigma \delta_l\;,
\hat{m}_{AB}=m_{AB},\;
\tilde{n}_{AB}=n_{AB}-\sigma\delta_l,\;
\hat{n}_{AB}=n_{AB},\;
m_1=\sigma\delta_l
\ee
in agreement with (\ref{1.14}). The other charges are unchanged. 
We denote the CNWF with the changed labels (\ref{1.15}) by 
$T^{A,l,\sigma}_{\gamma_1,m}$. 

The 
consideration for the other loops are similar: The factor 
$(h^l_{\alpha_B^1})^\sigma$ leads to the change of charges
\be \label{1.16}
\tilde{m}_{AB}=m_{AB},\;
\hat{m}_{AB}=m_{AB}-\sigma\delta_l,\;
\tilde{n}_{AB}=n_{AB},\;
\hat{n}_{AB}=n_{AB}+\sigma\delta_l,\;
m_1=\sigma\delta_l
\ee
We denote the CNWF with the changed labels (\ref{1.16}) by 
$T^{B,l,\sigma}_{\gamma_1,m}$. 

The factor 
$(h^l_{\alpha_B^2})^\sigma$ leads to the change of charges
\be \label{1.17}
\tilde{m}_{BC}=m_{BC}-\sigma \delta_l,\;
\hat{m}_{BC}=m_{BC},\;
\tilde{n}_{BC}=n_{BC}+\sigma\delta_l,\;
\hat{n}_{BC}=n_{BC},\;
m_2=\sigma\delta_l
\ee
We denote the CNWF with the changed labels (\ref{1.17}) by 
$T^{B,l,\sigma}_{\gamma_2,m}$. 

Finally, the factor 
$(h^l_{\alpha_C})^\sigma$ leads to the change of charges
\be \label{1.18}
\tilde{m}_{BC}=m_{BC},\;
\hat{m}_{BC}=m_{BC}+\sigma\delta_l,\;
\tilde{n}_{BC}=n_{BC},\;
\hat{n}_{BC}=n_{BC}-\sigma\delta_l,\;
m_2=\sigma\delta_l
\ee
We denote the CNWF with the changed labels (\ref{1.18}) by 
$T^{C,l,\sigma}_{\gamma_2,m}$. 

Note that still for $\ast\in\{AB,BC\}$
\be \label{1.19}
\tilde{m}_\ast+\tilde{n}_\ast
=\hat{m}_\ast+\hat{n}_\ast=N
\ee
We may summarise the relevant contributions to the action of the Hamiltonian 
constraint at vertices $A,B,C$ as 
\ba \label{1.20}
C_A \; T_{\gamma_0,m} &=& \sum_{l,\sigma}\;
\sigma\;
d^l_A(m_{ZA},n_{ZA},m_{AB},n_{AB},c_A)\; T^{A,l,\sigma}_{\gamma_1,m},\;
\nonumber\\
C_C \; T_{\gamma_0,m} &=& \sum_{l,\sigma}\;\sigma\;
d^l_C(m_{BC},n_{BC},m_{CD},n_{CD},c_C)\;\; T^{C,l,\sigma}_{\gamma_2,m},\;
\nonumber\\
C^I_B \; T_{\gamma_0,m} &=& \sum_{l,\sigma}\;\sigma\;
d_{B,I}^l(m_{AB},n_{AB},m_{BC},n_{BC},c_B)\;\; T^{B,l,\sigma}_{\gamma_I,m},\;
\ea
for $I=1,2$. 

Note that the labelling of the vector states 
$T^{A,l,\sigma}_{\gamma_1,m},T^{B,l',\sigma'}_{\gamma_1,m'}$ 
using labels A,B presents an overcounting, i.e. 
there are linear relations between them.
The same applies to 
$T^{C,l,\sigma}_{\gamma_2,m},T^{B,l',\sigma'}_{\gamma_2,m'}$
with respect to labels C,B.
This occurs precisely due to the phenomenon of non-unique parentage and will 
play a crucial role when solving the Hamiltonian constraint 
in the next subsections. On the other hand, the labels $(l,\sigma)$ 
taking six distinct possible values labelling the 
six distinct possible charges 
$\sigma\delta_l$ on the arcs and are not redundant.

\begin{figure}[hbt]
\includegraphics[width=18cm,height=12cm]{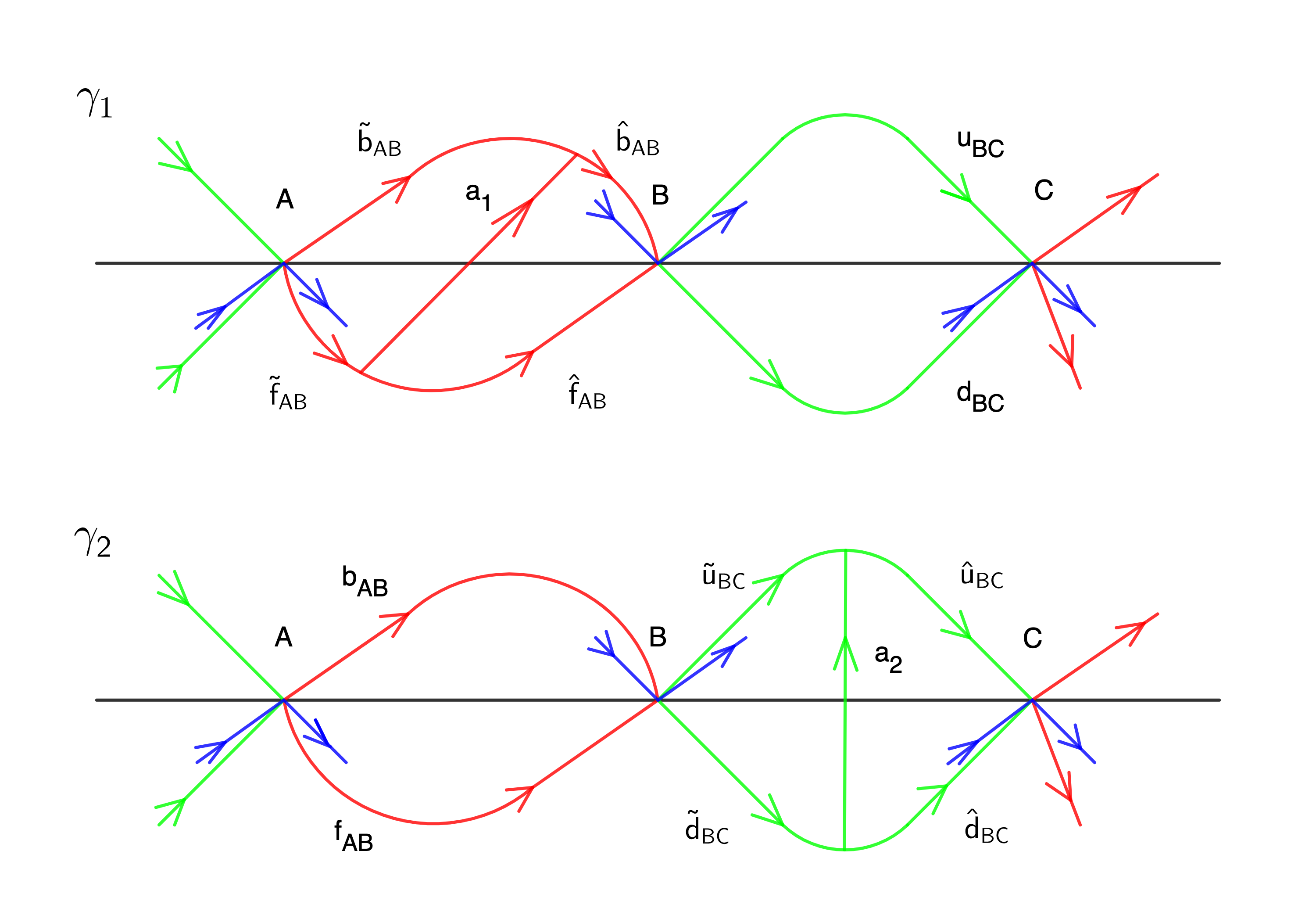}
\caption{The figure shows the two contributing children graphs.
Only the subdivided edges and arcs are highlighted, the rest 
of the graphs is identical to the parent graph.}
\end{figure}

Before closing this subsection, note that the children CNWF arising 
from the parent CNWF $T_{\gamma_0,m}$ can be such that 
one of the edges in the pairs 
$(\tilde{f}_{AB},\tilde{b}_{AB}),\;
(\hat{f}_{AB},\hat{b}_{AB})$ or 
$(\tilde{u}_{BC},\tilde{d}_{BC}),\;
(\hat{u}_{BC},\hat{d}_{BC})$, but not both if $N\not=0$,  
can become uncharged in the child 
by the action of the Hamiltonian constraint if 
the corresponding charge vector is of the form $\pm \delta_l$ in the parent. 
Since our solutions are by defintion such that 
all segments of $\gamma_0$ in the graphs $\gamma_1,\gamma_2$ are charged,
we do not need to worry about such solutions with `edge evaporation'.

\subsection{Statement of Quantum Einstein Equations for an example 
class of solutions}
\label{s1.6}

The Ansatz for an example solution to the Quantum Einstein Equations 
exhibiting propagation is 
the linear functional
\be \label{1.21}
\Psi_\Delta=\sum_{M\in \mathbb{Z}^6;\;I=1,2;\;l=1,2,3;\;\sigma=\pm 1}\; 
\kappa^{I,l,\sigma}_M\; \eta^{I,l,\sigma}_{M;\;\Delta}
\ee
The notation is as follows: Consider CNWF 
$T^{l,\sigma}_{\gamma_I,M;\Delta}$
over $\gamma_I$ where $\gamma_I$ are the child graphs described 
above and with fixed 
values of the data $\Delta:=(N,\{m_\ast\}_{\ast\in\;\{ZA,CD,DZ\}},
\{c_\ast\}_{\ast\in \{A,B,C\}})$ 
but with variable charges 
$M=(M_{AB}, M_{BC})\in \mathbb{Z}^6$ where for $I=1$
\be \label{1.22}
\tilde{m}_{AB}=M_{AB},\;\hat{m}_{AB}=M_{AB}-\sigma\delta_l, 
m_1=\sigma\delta_l, m_{BC}=M_{BC}
\ee
while for $I=2$
\be \label{1.23}
m_{AB}=M_{AB},\;\tilde{m}_{BC}=M_{BC}-\sigma\delta_l,\;
\hat{m}_{BC}=M_{BC},\;m_2=\sigma\delta_l
\ee
and of course $m_\ast+n_\ast=\hat{m}_\ast+\hat{n}_\ast=
\tilde{m}_\ast+\tilde{n}_\ast=N$ for all $\ast\in\{ZA,AB,BC,CD,DZ\}$.
Then $\eta^{I,l,\sigma}_{M;\Delta}$ is the diffeomorphism average
of $T^{l,\sigma}_{\gamma_I,M;\Delta}$ normalised such that 
that for any CNWF $T_{\gamma,m}$ we have 
\be \label{1.24}
\eta^{I,l,\sigma}_{M;\Delta}[T_{\gamma,m}]=
\delta_{[\gamma_I],[\gamma]}\;\langle T^{l,\sigma}_{\gamma_I,M;\Delta},\;
T_{\gamma_I,m} \rangle
\ee
where $\langle,.\rangle$ is the inner product on the kinematical 
LQG Hilbert space and $[\gamma]$ is the set 
$\{\varphi(\gamma),\;\varphi\in {\rm Diff}(\sigma)\}$ with Diff$(\sigma)$ 
denoting the group of semi-analytic diffeomorphisms of the spatial manifold 
$\sigma$ underlying the canonical formulation. Note that we dropped the 
vertex label $*\in \{A,B,C\}$ in $T^{*,l,\sigma}_{\gamma_I,M;\Delta}$
introduced in the previous subsection thus removing the overcounting.
In (\ref{1.24}) we exploited that the result vanishes unless the 
diffeomorphism classes of $\gamma,\gamma_I$ coincide.

In (\ref{1.24})
we have also exploited that 
$\gamma_0,\gamma_1,\gamma_2$ 
cannot have any graph symmetries \cite{GraphSymm} which would otherwise
unnecessarily complicate our discussion so that there is precisely one term 
in the distribution $\eta^{I,l,\sigma}_{M,\Delta}$ that contributes.
Here a graph symmetry of a graph $\gamma$ is  
semi-analytic diffeomorphism $\varphi$ that preserves $\gamma$ as a set 
but permutes its vertices and edges possibly including orientation reversal.
An orientation reversal for an edge $e$ of a CNWF over $\gamma$ is 
equivalent to a charge reflection $m_e\mapsto -m_e$ and a permutation of 
edges leads to a corresponding permutation of charges. The existence of graph
symmetries implies that CNWF over the same graph but different charges 
are in the same diffeomorphism orbit and thus their diffeomorphism 
averages are the same and thus (\ref{1.24}) would have to be 
corrected by a sum over the corresponding possible charge configurations 
that get identified under diffeomorphism averaging.    
   
To achieve a trivial graph symmetry group  
was the reason to add the extra structure $u_{ZA},d_{ZA},f_{CD},b_{CD},
l_A,l_B,l_C$ to $\gamma_0$: If we would drop these and connect $A,C$ by 
lines $u_{CA},d_{CA}$ similar to $f_{ZA},b_{ZA}$ then reflection 
in the y,z plane would exchange $f_{AB},b_{AB}$ and map $a_1$
to $a_1^{-1}$, reflection in the 
x,y plane would exchange $u_{BC},d_{BC}$ and map $a_2$ to 
$a_2^{-1}$ and and e.g. rotation by $\pi$ 
about the z-axis through $B$ followed by rotation by $\pi/2$ about the 
y-axis would map $(f_{AB},b_{AB},u_{BC},d_{BC})$ to 
$(u_{BC}^{-1},d_{BC}^{-1},f_{AB}^{-1},b_{AB}^{-1})$ 
and similar maps for $u_{CA}, d_{CA}$
leading to corresponding complicated 
identifications of charges in $\eta^{I,l,\sigma}_M$. With the extra structure
provided, the latter type of identifications and other similar ones 
are suppressed since a semi-analytic 
diffeomorphism cannot un-knot a loop.
Furthermore, while a semi-analytic diffeomorphism can be local,
it still has to be at least $C^1$. This prevents e.g. the existence of 
a local semi-analytic diffemorphism which preserves all parts of say
$\gamma_1$ while interchanging $f_{AB},b_{AB}$ for it would have to preserve
in particular $u_{BC}, d_{BC}, l_B$ at $B$. To see this in detail,
consider some parametrisation $[0,1]\ni t\mapsto s_e(t)$ where $s_e$ is an 
anylytic segment incident at and outgoing from $B$ respectively of the 
edge $e$ where $e$ is one of
$f_{AB}^{-1},b_{AB}^{-1},u_{BC},d_{BC}$ or the beginning segment of $l_B$.
Suppose there exists a semi-analytic diffeomorphism $\varphi$ with  
\ba \label{1.25a}
&& \varphi(s_{f_{AB}^{-1}}(t))=b_{AB}^{-1}(f_1(t)),\;
\varphi(s_{b_{AB}^{-1}}(t))=f_{AB}^{-1}(f_2(t)),\;
\varphi(s_{u_{BC}}(t))=u_{BC}(f_3(t)),\;
\nonumber\\
&& \varphi(s_{d_{BC}}(t))=d_{BC}(f_4(t)),\;
\varphi(s_{l_B}(t))=l_B(f_5(t)),\;
\ea
where $f_I$ are reparametrisations of $[0,1]$, in particular $f_I(0)=0,\;
df_I/dt>0$. Let $M$ be the matrix $(\partial \varphi/\partial x)(B)$ and 
$\lambda_I=(df_I/dt)(0)>0$. Taking the derivatives of (\ref{1.25a}) at $t=0$
gives  
\ba \label{1.25b}
&& M(b_1-b_2)=\lambda_1(-b_1-b_2),\;
M(-b_1-b_2)=\lambda_2(b_1-b_2),\;
M(b_3+b_2)=\lambda_3(b_3+b_2),\;
\nonumber\\
&& M(-b_3+b_2)=\lambda_4(-b_3+b_2),\;
M(-b_1+b_2)=\lambda_5(-b_1+b_2)
\ea
where $b_j$ denotes the Cartesian basis of $\mathbb{R}^3$. The 
first and fifth equation are in contradiction leading to 
$\lambda_1=\lambda_5=0$. Without the fifth equation,
the first four equations would be satisfied for $\lambda_I=1$ and $M$ the 
reflection at the y,z plane. Likewise, a simultaneous interchange of 
$f_{AB}, b_{AB}$ and $u_{BC}, d_{BC}$ while preserving $l_B$ would lead to
a contradiction between the first and fifth equation. An interchange 
of $u_{BC}, d_{BC}$ preserving all other parts of the graph, in 
particular also the end segment of $l_B$, would lead to
\ba \label{1.25c}
&& M(b_1-b_2)=\lambda_1(b_1-b_2),\;
M(-b_1-b_2)=\lambda_2(-b_1-b_2),\;
M(b_3+b_2)=\lambda_3(-b_3+b_2),\;
\nonumber\\
&& M(-b_3+b_2)=\lambda_4(b_3+b_2),\;
M(-b_3+b_2)=\lambda_5(-b_3+b_2)
\ea
and now the fourth and fifth equation are in contradiction, leading to 
$\lambda_4=\lambda_5=0$. Without the fifth equation the first four equations
can be satisfied by $\lambda_I=1$ and $M$ the reflection at the x,y plane. 
Finally a mapping between the pairs $(f_{AB},u_{AB})$ and $(u_{BC}, d_{BC})$ 
possibly with interchanges within a pair and possibly with 
orientation reversal would need to map the whole structure to the one 
with the sequence of vertices reveresed preserving $B$, in particular 
$l_A\mapsto l_C^{\pm 1},\; l_B\mapsto l_B^{\pm 1},\;  
l_C\mapsto l_A^{\pm 1}$ but this cannot be a graph symmetry because 
$l_A,l_C$ have different topology. 

Even if $l_A, l_C$ would have the same topology so that a graph symmetry 
inducing such a pair exchange 
would be conceivable, a
corresponding identification of charges on these edges possibly including
reflections would be accompanied by an associated identification of 
charges between the pairs of edges $(u_{ZA},d_{ZA})$ and $(f_{CD}, b_{CD})$
respectively. Since we have fixed the data $\Delta$, if we
pick the pair of charges $(m_{ZA},n_{ZA})$ and $(m_{CD}, n_{CD})$ respectively 
such that there is no combination of permutation maps or single entry 
reflection maps between them, then at most one of these two CNWF is 
not annihilated by $\eta^{I,l,\sigma}_{M;\Delta}$. The same 
can be achieved by 
picking the charges on the loops such that there are no reflection or 
permutation maps between them. Thus even in this case (\ref{1.24}) remains 
valid.\\
\\
The $\kappa^{I,l,\sigma}_M$ are complex coefficients that are to be 
determined by asking 
that $\Psi$ obeys the Quantum Einstein Equations, that is
\be \label{1.25}
\Psi_\Delta[(U(\varphi)-{\rm id})\;\; T_{\gamma,m}]=0=
\Psi_\Delta[C(f)\; T_{\gamma,m}]\;\; 
\forall\;\;\varphi\in \;{\rm Diff}(\sigma),\;
f,\;\gamma,\;m,\;\;
\ee
Here $U$ is the unitary representation of Diff$(\sigma)$ densely defined by 
\be \label{1.26}
U(\varphi)\;T_{\gamma,\{m_e\}}=T_{\varphi(\gamma),\{m_{\varphi(e)}=m_e\}}
\ee
from which we see that the first condition in (\ref{1.25}) is already 
satisfied. The $T_{\gamma,m}$ are w.l.g. solutions to the Gauss constraint
(otherwise (\ref{1.25}) is again trivially solved). 

Moreover, 
$\kappa^{I,l,\sigma}_M$ is set to zero for those $M$ for which 
not all edges of $\gamma_I$ are charged in 
$T^{l,\sigma}_{\gamma_I,M;\Delta}$ according to (\ref{1.22}) and 
(\ref{1.23}). Thus we impose on the data $\Delta$ that $N\not=0,
m_\ast\not=0,N; \; \ast\in \{ZA,CD,DZ\}$ and that for 
$I=1$ and each $l,\sigma$ the coefficient $\kappa^{I,l,\sigma}_M$ vanishes 
when
\be \label{1.26a}
M_{AB}\in \{0,N,-\sigma\delta_l,N-\sigma\delta_l\},\;{\rm or}\;\;
M_{BC}\in\{0,N\}
\ee
and for 
$I=2$ and each $l,\sigma$ the coefficient $\kappa^{I,l,\sigma}_M$ vanishes 
when
\be \label{1.26b}
M_{AB}\in\{0,N\},\;\; {\rm or}\;\;
M_{BC}\in\{0,N,\sigma\delta_l,N+\sigma\delta_l\}
\ee
This avoids that there can be child graphs with ``erased edges'' 
in addition to the 
ones considered. Note 
that (\ref{1.26a}) and (\ref{1.26b}) would not be 
necessary for the closable operator of section \ref{s1.3a}
because by construction the closable operator does not have SNWF in its image 
for which not all edges of the parent graph are charged. In what 
follows we consider the non closable operator for simplicity. Also note 
that child graphs with ``erased edges'' need to be carefully taken into 
account among the first generation graphs \cite{QSDII} when constructing
higher generation solutions. 

To solve the Hamiltonian constraint we note that due to our specific
Ansatz 
$\Psi_\Delta[C(f)\;T_{\gamma,m}]\equiv 0$
is automatically satisfied unless the child graphs of $\gamma$ that appear
in the CNWF decomposition of $C(f)\; T_{\gamma,n}$ lie in the diffeomorphism 
class of either $\gamma_1$ or $\gamma_2$. This means that $\gamma\in 
[\gamma_0]$
because the Hamiltonian action consists in adding arcs between pairs of edges
meeting in at least tri-valent vertices. For $\gamma\in [\gamma_0]$ we may 
choose w.l.g. the representative $\gamma=\gamma_0$ and 
we can choose the support of $f$ solely in mutually disjoint 
neighbourhoods of the vertices $A,B,C$ 
which means that we obtain the following system of linear equations
\be \label{1.27}
\Psi_\Delta[C_A\; T_{\gamma_0,m}]=0,\;\;
\Psi_\Delta[(C_B^1+C_B^2)\;T_{\gamma_0,m}]=0,\;\;
\Psi_\Delta[C_C\; T_{\gamma_0,m}]=0
\ee
for all $m_\ast\in \mathbb{Z}^3,\;\ast\in \;\{ZA, AB, BC, CD, DZ\}$
and $c_\ast\in \mathbb{Z}^3,\; \ast \in \{A,B,C\}$. 
Again (\ref{1.27}) is trivially satisfied unless   
$(N:=m_{ZA}+n_{ZA},\{m_\ast\}_{\ast\in\;\{ZA,CD,DZ\}},c_{\ast\in \{A,B,C\}})
=\Delta$. To evaluate (\ref{1.27}) 
for such  CNWF we notice 
\be \label{1.28}
T^{l,\sigma}_{\gamma_I,M;\Delta}=T^{B,l,\sigma}_{\gamma_I,m}
\ee
where it is understood that the decomposition of the data contained in $m$ 
into $(m_{AB},m_{BC})$ and the rest matches the data $M$ and $\Delta$ 
respectively. Thus 
\ba \label{1.29}
&& \langle T^{l,\sigma}_{\gamma_I,M;\Delta},\;
T^{B,l',\sigma'}_{\gamma_J,m} \rangle=
\delta_{I,J}\;\delta_{l,l'}\;
\delta_{\sigma,\sigma'}\;
\delta_{M_{AB},m_{AB}}\;
\delta_{M_{BC},m_{BC}}\;
\nonumber\\
&& \langle T^{l,\sigma}_{\gamma_1,M;\Delta},\;
T^{A,l',\sigma'}_{\gamma_1,m} \rangle=
\delta_{l,l'}\;\delta_{\sigma,\sigma'}\;
\delta_{M_{AB},m_{AB}+\sigma\delta_l}\;
\delta_{M_{BC},m_{BC}}\;
\nonumber\\
&& \langle T^{l,\sigma}_{\gamma_2,M;\Delta},\;
T^{C,l',\sigma'}_{\gamma_2,m} \rangle
=\delta_{l,l'}\;
\delta_{\sigma,\sigma'}\;
\delta_{M_{AB},m_{AB}}\;
\delta_{M_{BC},m_{BC}+\sigma\delta_l}
\nonumber\\
&& \langle T^{l,\sigma}_{\gamma_1,M;\Delta},\;
T^{C,l',\sigma'}_{\gamma_2,m} \rangle
= \langle T^{l,\sigma}_{\gamma_2,M;\Delta},\;
T^{A,l',\sigma'}_{\gamma_1,m,N} \rangle
=0
\ea
We thus obtain the following system of non-trivial equations 
\ba \label{1.30}
0 &=& \Psi_\Delta[C_A\; T_{\gamma_0,m}]=\sum_{l,\sigma}\;\sigma\;
\kappa^{1,l,\sigma}_{M_{AB}=m_{AB}+\sigma\delta_l,M_{BC}=m_{BC}}
\;\;d^l_A(m_{ZA},n_{ZA},m_{AB},n_{AB},c_A)\; 
\nonumber\\
0 &=& \Psi_\Delta[C_C\; T_{\gamma_0,m}]=\sum_{l,\sigma}\;\sigma\;
\kappa^{2,l,\sigma}_{M_{AB}=m_{AB},M_{BC}=m_{BC}+\sigma \delta_l}
\;\;d^l_C(m_{BC},n_{BC},m_{CD},n_{CD},c_C)
\nonumber\\
0 &=& \Psi_\Delta[(C^1_B+C_B^2)\; T_{\gamma_0,m}]=
\sum_{I,l,\sigma}\;\sigma\;
\kappa^{I,l,\sigma}_{M_{AB}=m_{AB},M_{BC}=m_{BC}}
\;\; d_{B,I}^l(m_{AB},n_{AB},m_{BC},n_{BC},c_B)
\ea
to be solved for all $(m_{AB},m_{BC})\in \mathbb{Z}^6$.

\subsection{Solving the example class and discussion}
\label{s1.7}

{\bf Preparation}\\
\\
The structure of the solutions of the system (\ref{1.30}) of course depends
sensitively on the coefficients $d^l_A,\; d^l_C, d^l_{B,I}$ with
$I=1,2$. We note 
that $d^l_A$ depends on $m_{AB};m_{ZA},c_A,N$,   
that $d^l_B$ depends on $m_{BC};m_{CD},c_B,N$ and 
that $d^l_{B,I}$ depends on $m_{AB};m_{BC},c_B,N$.
The dependence on $N$ is via $m_\ast+n_\ast=N$.

To simplify the subsequent discussion, we will drop the dependence 
of the these functions on the fixed structure $\Delta$ and introduce the 
notation $p:=m_{AB}, \; q:= m_{BC}$ in order not to clutter the formulae.
Furthermore, following the discussion in section \ref{s1.4a} we assume
that $\Delta$ has been chosen such  
that the number of configurations $p,q$ for which one or several of the 
$\vec{d}_A, \vec{d}_C,\; \vec{d}_{B,1},\;\vec{d}_{B,2}$ or that 
$\vec{d}_{B,1},\;\vec{d}_{B,2}$ are linearly dependent is a finite 
subset of $\mathbb{Z}^3$ or $\mathbb{Z}^6$ respectively.   
\\
\\
{\bf Existence and explicit construction of Solutions}\\
\\
We note that we have coefficients $\kappa^{I,l,\sigma}_M$ that 
depend on indices $I\in \{1,2\},\; l\in \{1,2,3\},\;\sigma\in \{+1,-1\}$ and 
$M=(p,q)\in\mathbb{Z}^6$, that is, 
12 coefficients per point $M\in \mathbb{Z}^6$,
except for configurations $M$ depending on $(I,l,\sigma)$
which are excluded by (\ref{1.26a}) and (\ref{1.26b}). 
These coefficients are subject to the 3 constraints (\ref{1.30}) for every 
point
$M\in \mathbb{Z}^6$ except for those 
such that $C(f)\;T_{\gamma_0,m}$ has an expansion 
into CNWF over $\gamma_1,\gamma_2$ such that for all appearing CNWF
not all edges of $\gamma_1,\gamma_2$ are charged. 
Thus, except for those exceptional points, we have roughly counting 
12 times a $\mathbb{Z}^6$ worth of complex variables and 3 times 
a $\mathbb{Z}^6$ worth of equations and thus expect that the system 
(\ref{1.30}) admits a rich (in fact infinite) number of solutions. 
However not all of them may display propagation. To see this, let us
refer to the third equations in (\ref{1.30}) as the ``B-equations'', and the 
first and second equations in (\ref{1.30}) as the ``A-equations'' 
and ``C-equations'' respectively. Consider the subspace of solutions 
derived by setting
$\kappa^{2,l,\sigma}=0$
and call the remaining equations the 1-equations. 
In a similar manner we define the 
2-equations by setting the coefficients $\kappa^{2,l,\sigma}=0$.
Note that the 1-equations respectively 2-equations only involve the 
`$\kappa$' coefficients 
for the child of type 1 and 2 respectively.
Next, consider the 1-equations. The C equations are trivially satisfied and 
we are left with the A-equations and the B-equations that now impose conditions
only on $\kappa^{1,l,\sigma}$. Likewise, for  
the 2-equations, the A equations are trivially satisfied and 
we are left with the C-equations and the B-equations that now impose conditions
only on $\kappa^{2,l,\sigma}$. In both cases, we obtain 
2$\mathbb{Z}^6$ equations for the corresponding 
6$\mathbb{Z}^6$ left over 
of $\kappa$ 
coefficients which should yield
4$\mathbb{Z}^6$ solutions to these equations and hence 
4$\mathbb{Z}^6$ solutions to all $A,B, C$ equations.
Altogether we should thus obtain 8$\mathbb{Z}^6$ solutions 
which are not propagating because each of them involves only a single child 
graph. This leaves us with a 9$\mathbb{Z}^6$  - 8$\mathbb{Z}^6$  
i.e a $\mathbb{Z}^6$  worth of solutions.

This expectation is indeed borne out by the detailed solution construction 
method below so that  
the propagation degree of the solution (to be defined below)
class is only unity, i.e. only 
one complex variable can be accounted for propation rather than nine.   

We proceed as follows. Turning to the B-equations, we note that they are
ultra-local with respect to both p,q and and can be 
written more explicitly as
\be \label{1.31}
\vec{\Delta}^{1}_{p,q}\cdot\vec{d}_{B,1}(p,q)+
\vec{\Delta}^{2}_{p,q}\cdot\vec{d}_{B,2}(p,q)=0
\ee
where we have introduced for $I=1,2$ 
\be \label{1.32}
(\vec{\Delta}^I_{p,q})^l:=
\kappa^{I,l,+}_{p,q}-\kappa^{I,l,-}_{p,q}
\ee
It can therefore be solved algebraically for each $p,q$ separately and we drop
the dependence on $p,q$ as far as the B-equations are concerned. \\
Case 0: Both $\vec{d}_{B,I},\; I=1,2$ vanish.\\
Let $\vec{d}_a,\; a=1,2,3$ be the Cartesian basis of $\mathbb{R}^3$. Then 
\be \label{1.32a}
\vec{\Delta}^I=\gamma^I\; \vec{d}_3+\alpha^I\; \vec{d}_1+\beta^I\; \vec{d}_2
\ee
is the general solution with arbitrary 6 parameters $\alpha^I,\beta^I,
\gamma^I;\; I=1,2$. We arbitrarily set $\beta^2=\alpha^1:=\nu$ thus 
deliberatively reducing it to five parameters although this is not 
necessary.\\
Case 1a: Only $\vec{d}_{B,2}$ vanishes.\\
Let $\vec{d}_1:=\vec{d}_B^1/||\vec{d}_B^1||$ 
and $\vec{d}^2,\vec{d}^3$ be any two vectors such that $\vec{d}_a,\; a=1,2,3$
is an orthonormal basis of $\mathbb{R}^3$. Then
\be \label{1.32b}
\vec{\Delta}^1=\gamma^1\; \vec{d}_3+\beta^1\; \vec{d}_2,\;\;
\vec{\Delta}^2=\gamma^2\; \vec{d}_3+\alpha^2\; \vec{d}_1+\beta^2\; \vec{d}_2
\ee
is the general solution with arbitrary 5 parameters $\gamma^1,\gamma^2,
\alpha^2,\beta^1,\nu:=\beta^2$.\\
Case 1b: Only $\vec{d}_{B,1}$ vanishes.\\
Let $\vec{d}_2:=\vec{d}_B^2/||\vec{d}_B^2||$ 
and $\vec{d}^1,\vec{d}^3$ be any two vectors such that $\vec{d}_a,\; a=1,2,3$
is an orthonormal basis of $\mathbb{R}^3$. Then
\be \label{1.32bb}
\vec{\Delta}^1=\gamma^1\; \vec{d}_3+\alpha^1\; \vec{d}_1+\beta^1\; 
\vec{d}_2,\;\;
\vec{\Delta}^2=\gamma^2\; \vec{d}_3+\alpha^2\; \vec{d}_1
\ee
is the general solution with arbitrary 5 parameters $\gamma^1,\gamma^2,
\alpha^2,\beta^1,\nu:=\alpha^1$.\\
Case 1c: $\vec{d}_{B,I},\;I=1,2$ are co-linear but non-vanishing.\\
Thus there is $\mu\not=0$ such that $\vec{d}_{B,2}=\mu\;
\vec{d}_{B,1}$. Let $\vec{d}_1:=\vec{d}_B^1/||\vec{d}_B^1||$ 
and $\vec{d}^2,\vec{d}^3$ be any two vectors such that $\vec{d}_a,\; a=1,2,3$
is an orthonormal basis of $\mathbb{R}^3$. Then 
\be \label{1.32c}
\vec{\Delta}^1=\gamma^1\; \vec{d}_3+(-\mu\nu)\; 
\vec{d}_1+\beta^1\; \vec{d}_2,\;\;
\vec{\Delta}^2=\gamma^2\; \vec{d}_3+\nu\; \vec{d}_1+\alpha^2\;\vec{d}_2
\ee
is the general solution with arbitrary 5 parameters $\gamma^1,\gamma^2,
\alpha^2,\beta^1,\nu$.\\
Case 2: $\vec{d}_{B,I},\;I=1,2$ are linearly independent.\\
We write with $\vec{d}_{B,3}:=
\vec{d}_{B,1}\times \vec{d}_{B,2}$ and 
expand
\be \label{1.33}
\vec{\Delta}^I=
\gamma^I\;\vec{d}_{B,3}
+\alpha^I\;\vec{d}_{B,1}
+\beta^I\;\vec{d}_{B,2}
\ee
It follows
\be \label{1.34}
||\vec{d}_{B,1}||^2\; \alpha^1
+||\vec{d}_{B,2}||^2\; \beta^2
+(\vec{d}_{B,1}\cdot \vec{d}_{B,2})\;(\alpha^2+\beta^1)=0
\ee 
Using $||\vec{d}_{B,1}||^4+||\vec{d}_{B,2}||^4>0$ 
we find the general solution
\be \label{1.35}
\alpha^1=||\vec{d}_{B,1}||^2\;\mu+   
||\vec{d}_{B,2}||^2\;\nu,\;\;
\beta^2=||\vec{d}_{B,2}||^2\;\mu-   
||\vec{d}_{B,1}||^2\;\nu,\;\;
\mu=-
\frac{(\vec{d}_{B,1}\cdot \vec{d}_{B,2})\;
(\alpha^2+\beta^1)}{||\vec{d}_{B,1}||^4+||\vec{d}_{B,2}||^4}
\ee
depending on 5 parameters $\gamma^1,\gamma^2,\alpha^2,\beta^1,\nu$.\\
\\
To summarise: Whatever the coefficients $\vec{d}_{B,I}$, we find an at
least 5 parameter set of solutions to all B-equations. 
Note that the case 2 that 
$\vec{d}_{B,1}(p,q),\vec{d}_{B,2}(p,q)$ are non-vanishing and linearly 
independent is satisfied at generic points $p,q$ as argued 
in section \ref{s1.4a}. There is {\it necessarily}
a non-trivial coupling of $\vec{\Delta}^1,\vec{\Delta}^2$ in cases 
1c and 2.

Turning to the A-equations and C-equations, 
we consider the equivalent set of coefficient functions $\Delta^{I,l}$ and 
$\kappa^{I,l,-}$ where $\kappa^{I,l,+}=\Delta^{I,l}+\kappa^{I,l,-}$ and 
$\Delta^{I,l}$ is now parametrised by a linear combination 
of the five free functions $\gamma^I,\nu,\alpha^2,\beta^1$. Accordingly, 
we write the A-equations and C-equations in (\ref{1.30}) as 
\ba \label{1.36}
0 &=&
\sum_l\;
[\Delta^{1,l}_{p-\delta_l,q}+
\kappa^{1,l,-}_{p-\delta_l,q}-
\kappa^{1,l,-}_{p+\delta_l,q}] \; d^l_A(p,q)
\nonumber\\
0 &=&
\sum_l\;
[\Delta^{2,l}_{p,q+\delta_l}+
\kappa^{2,l,-}_{p,q+\delta_l}-
\kappa^{2,l,-}_{p,q-\delta_l}] \; d^l_C(p,q)
\ea
We note that the first of these two equations is ultra-local in q and 
the second in $p$. Moreover, $\kappa^{1,l,-},\gamma^1$ are only involved 
in the first equation, $\kappa^{2,l,-}, \gamma^2$ only in the second
while generically both depend on $\alpha^2,\beta^1,\nu$ through which they are 
coupled in Case 2 (which is expected to be the generic case, see section \ref{s1.4a}). It is therefore natural to solve the first and second equation 
respectively in terms of $\kappa^{1,l,-},\gamma^1$ and 
$\kappa^{2,l,-},\gamma^2$ respectively. We will show that it is 
possible to solve both equations just in terms of 
$\kappa^{1,l,-},\;\kappa^{2,l,-}$ respectively,
thus leaving $\gamma^1,\gamma^2$ untouched.

Let us abbreviate 
\be \label{1.37a}
\rho_A^l=\kappa^{1,l,-},\;
\rho_C^l=\kappa^{2,l,-},\;
u_A^l(p,q):=\Delta^{1,l}_{p+\delta_l,q},\;
u_C^l(p,q):=-\Delta^{2,l}_{p,q+\delta_l},\;
\ee
and define the discrete derivatives
\be \label{1.38}
(\partial^A_l g)(p,q):=g(p+\delta_l,q)-g(p-\delta_l,q),\;
(\partial^C_l g)(p,q):=g(p,q+\delta_l)-g(p,q-\delta_l)
\ee
to cast (\ref{1.36}) into the form (at each $(p,q)$)
\be \label{1.39}
\sum_l\; d_A^l\; [\partial^A_l \rho_A^l-u_A^l]=0,\;
\sum_l\; d_C^l\; [\partial^C_l \rho_C^l-u_C^l]=0
\ee
As we wish to keep $u_A^l, u_C^l$ unconstrained, we consider these as 
inhomogeneities in the linear discrete PDE problems (\ref{1.39}) for the 
three functions $\rho_A^l$ and $\rho_C^l$ respectively. Due 
to linearity, the general solution $\rho_A^l$ 
is a linear combination of a particular
solution to the inhomogeneous problem $\rho_{A,{\rm inhom}}^l$
and of the general solution to the 
homogeneous problem $\rho_{A,{\rm hom}}^l$. If we had only one function each  
$\rho_A^l:=\rho_A,\rho_C^l:=\rho_C$ available
then the construction of the solution to the inhomogeneous problem would
read
\be \label{1.39a}
\sum_l\; d_A^l\; \partial^A_l \rho_A=f_A,\;
\sum_l\; d_C^l\; \partial^C_l \rho_C=f_C
\ee
where 
\be \label{1.37}
f_A(p,q):=   
\sum_l\;\Delta^{1,l}_{p+\delta_l,q}\; d^l_A(p,q),\;\;
f_C(p,q):=-   
\sum_l\;\Delta^{2,l}_{p,q+\delta_l}\; d^l_C(p,q),\;\;
\ee
This is the the discrete version of two independent PDE's 
of the form 
$\vec{d}\cdot\nabla \rho=f$ with a vector field $\vec{d}$.
One would then try to solve the discrete 
PDE's by a discrete version of the characteristic method \cite{PDE} by 
inverting the logic of solving a PDE numerically using finite difference 
techniques. As is well known from the continuous version, this can 
lead to global existence questions depending on the details of $\vec{d}$.

Fortunately, as we have {\it three} functions rather than just 
one available, an inhomogeneous 
solution is readily provided explicitly: We simply solve for each $l$
separately
\be \label{1.39b}
\partial^A_l \rho_A^l=u_A^l,\;\;
\partial^C_l \rho_C^l=u_C^l
\ee
These can be solved explicity by integration if we prescribe 
``initial values''.
We will display this for the $l=1$ direction, the $l=2,3$ directions 
can be treated by cyclic permutation of the roles of $l=1,2,3$. The 
integration of (\ref{1.39b}) for $l=1$ requires initial data
$\tau^1_{A,0}(p_2,p_3;q),\tau^1_{A,1}(p_2,p_3;q)$
for $\rho_A^1(p,q)$ at $p_1=0,1$ and 
$\tau^1_{C,0}(p;q_2,q_3),\tau^1_{C,q}(p;q_2,q_3)$ for 
$\rho^1_C(p,q)$ at $q_1=0,1$. One finds (we drop dependence on 
ultra local variables and on $2,3$ components of remaining
charge vectors involved) 
\ba \label{1.42}
\rho^1_A(2p_1) 
&=& \sum_{k=0}^{p_1-1} 
u_A^1((p)_1=2k+1)
+\tau^1_{A,0},\;p_1>0
\nonumber\\
\rho_A^1(2p_1+1) 
&=& \sum_{k=1}^{p_1} 
u_A^1((p)_1=2k)+\tau^1_{A,1},\;p_1>0
\nonumber\\
\rho^1_A(-2p_1) 
&=& 
-\sum_{k=0}^{p_1-1} u_A^1((p)_1=-(2k+1))+\tau^1_{A,0},\;p_1>0
\nonumber\\
\rho_A^1(-(2p_1+1)) 
&=& -\sum_{k=0}^{p_1} u_A^1((p)_1=-2k)+\tau^1_{A,1},\;p_1\ge 0
\ea
and similar for $\rho_C^1,\;\tau^1_{C,0},\;\tau^1_{C,1}$ 
with $p_1$ substituted by $q_1$.

This proves existence of an inhomogeneous solution with no restriction 
on $\vec{\Delta}^I$ other than that coming from the B-equations 
which were already solved. We are free to add 
a solution of the homogeneous equation 
\be \label{1.39c}
\sum_l\; d_A^l\; \partial^A_l \rho_A^l=0,\;
\sum_l\; d_C^l\; \partial^C_l \rho_C^l=0
\ee
In contrast to the inhomogeneous solution just provided, the general solution 
of (\ref{1.39c}) will depend on the discrete ``characteristics'' (in the 
sense of PDE theory) determined by the discrete ``vector fields'' 
$\vec{d}_A,\vec{d}_C$. To see how one would proceed,
consider the A-equation, start at $p=0$ 
and suppose that the component $d^1_A(0)$ is 
not vanishing. Then, given an initial datum at $p=0,q$ for $\rho_A^1$, 
we can solve $\rho_A^1$ along an ``$l=1$ flow line'', in terms of 
$\rho_A^2,\rho_A^3$ by (\ref{1.42}) if we simply redefine 
$u_A^1:=\frac{1}{d_A^1}\;\sum_{l=2,3}\; d_A^l\; \partial^A_l \rho_A^l$.
This works until $d_A^1=0$. When that happens, but $d_A^2\not=0$ 
we may similarly solve $\rho_A^2$ in terms $\rho_A^3$ 
along an ``$l=2$ flow line'' as long as $d_A^2\not=0$ 
until $d_A^1\not=0$ again or both $d_A^1=d_A^2=0$. In the first case, we 
solve again along the $l=1$ line, in the second, if $d_A^3\not=0$ we 
keep $\rho_A^3$ constant along the $l=3$ line until either 
$d_A^1\not=0$ again in which case we proceed propagating along $l=1$, or 
$d_A^1=0,d_A^2\not=0$ again in which case proceed 
propagating along $l=2$ or $\vec{d}_A=0$. At the zeroes of the vector 
field there are obviously no conditions on the $\rho_A^l$ and we must supply
additional initial data there in order to extend the solution beyond. 
This way one finds the solution along the ``lexicographic 
integral curves'' of 
$\vec{d}_A$ (i.e. along directions in charge space of its 
non-vanishing components ordered lexicographically) 
which is possible until they cross which leads to ``shocks''.
We will not go into further details here but rather remark that as argued in 
section \ref{s1.4a}
i.  the vector fields $\vec{d}_A,\;\vec{d}_C$ and even 
their separate components have zeroes at 
most at isolated points and ii. we may 
always solve the homogeneous equation by the trivial solution. Then 
still the set of solutions to all constraints is infinite consisting 
of the five free parameters at each $(p,q)\in \mathbb{Z}^6$ that are 
left over after solving the B-equations and the six parameters parametrising
the initial data on two $\mathbb{Z}^5$ surfaces, that is 
$\tau^l_{A,0},\; \tau^l_{A,1}, \tau^l_{C,0},\;\tau^l_{C,1};\; l=1,2,3$.

There is one remaining subtlety: the equations (\ref{1.30}) were 
derived under the assumption that (\ref{1.26a}) and (\ref{1.26b}) hold, 
i.e. that $\kappa^{1,l,\sigma}(p,q)$ vanishes whenever 
$p\in \{0,N,-\sigma\delta_l,N-\sigma\delta_l\}$ or $q\in \{0,N\}$ and 
that $\kappa^{2,l,\sigma}(p,q)$ vanishes whenever 
$q\in \{0,N,\sigma\delta_l,N+\sigma\delta_l\}$ or $p\in \{0,N\}$.
Since $\Delta^{I,l}=\kappa^{I,l,+}-\kappa^{I,l,-}$, we 
have that $\kappa^{I,l,+}=\Delta^{I,l}+\kappa^{I,l,-}$. 
This in fact couples the B-equations that relate $\Delta^{1,l},\;\Delta^{2,l}$ 
with the A-equations on $\kappa^{1,l,-}$ and with the C-equations 
on $\kappa^{2,l,-}$. While this provides an additional
propagation effect, it jeopardises the derivation above which 
rested on the assumption that we can consider the 
$u_A^l\propto\Delta^{1,l},\;u_C^l\propto \Delta^{2,l}$
as independent of $\rho_A^l,\rho_C^l$. In order to avoid this we impose 
the stronger, uniform condition that $\kappa^{I,l,\sigma}$ vanishes 
whenever $p$ or $q$ lie in the set of six distinct (assuming 
$|\vec{N}_k|>2$ for  $k=1,2,3$) points 
$s_l:=\{0,\delta_l,-\delta_l,N,N+\delta_l,N-\delta_l\}$. That way, 
the vanishing condition on $\kappa^{I,l,\sigma}$ no longer depends on the 
labels $I,\sigma$ and now the vanishing of 
$\kappa^{I,l,+},\;\kappa^{I,l,-}$ on 
$S_l:=s_l\times \mathbb{Z}^3\cup \mathbb{Z}^3 \times s_l$ is equivalent to 
the vanishing of $\Delta^{I,l},\; \kappa^{I,l,-}$ on $S_l$.
To statisfy this stronger condition, we proceed as follows:
The B-equations are ultra-local, thus we simply set all five 
free coefficients that parametrise $\Delta^{I,l}$ to zero on $S_l$ because 
$\Delta^{I,l}$ is homogeneous linear in those. This in fact implies that 
all $\Delta^{I,l'}(p,q),\; I=1,2,\; l'=1,2,3$ vanish at any 
$(p,q)\in S_l$, not only the $l'=l$ components.  
Considering the A-equations
and if we set a possible homogeneous solution to zero, 
we must ensure that the solutions $\rho_A^l,\rho_C^l$ vanish on $S_l$. 
Thus explicitly the solution $\rho_A^1$ given in 
(\ref{1.42}) must vanish on $S_1$. Recall that 
$\rho_A^1$ depends on initial data 
$\tau^1_{A,0}(p_2,p_3;q),\;\tau^1_{A,1}(p_2,p_3;q)$ on the hypersurfaces 
$\Sigma^1_0=\{(p_1=0,p_2,p_3;q);\;p_2,p_3\in \mathbb{Z}, q\in \mathbb{Z}^3\}$
and
$\Sigma^1_1=\{(p_1=1,p_2,p_3;q);\;p_2,p_3\in \mathbb{Z}, q\in \mathbb{Z}^3\}$.
Now $s_1=\{(\{0,\pm 1\},0,0),(N_1+\{0,\pm 1\},N_2,N_3)\}$ thus 
the points in $s_l\times \mathbb{Z}^3$ with $(p_2,p_3)=(0,0)$ and 
$(p_2,p_3)=(N_2,N_3)$ lie on different $l=1$ lines.
It is then straightforward to see that 
we can guarantee 
the vanishing condition on $\rho_A^1$ at those points by simply restricting 
one of the initial data in (\ref{1.42}) on each of the two distinct  $l=1$ lines. 
On the other hand the points 
in $\mathbb{Z}^3 \times s_1$ fill out all of $p$ space at six discrete 
values of $q\in s_1$. However, by (\ref{1.37a}) and by the stronger 
assumption on $\Delta^{I,l}$ also the function $u_A^1$ vanishes there 
identically. We may therefore consistently set $\rho_A^1\equiv 0$ on 
$\mathbb{Z}^3\times s_1$. 

Summarising, for $q\in s_1$ we set 
$\rho_A^1\equiv 0$ for all $p\in \mathbb{Z}^3$. For $p\in s_1,\; 
q\in\mathbb{Z}^3-s_1$ we can grant $\rho_A^1=0$ by choosing one of the 
initial data. Similar considerations hold for $\rho_C^1$ and the other 
directions $l=2,3$. Note that if we would add a homogeneous 
solution we could satisfy the vanishing conditions by similar restrictions.
Finally note that the sets $S_l$ are of  ``measure zero'' in $\mathbb{Z}^6$
having ``codimension'' 3. \\
\\
{\bf Non-normalisability and further properties of the solutions}\\
\\
The solution (\ref{1.42}) shows the following features:\\
1.\\
Even if the free data 
$\gamma^1,\nu,\alpha^2,\beta^1$ 
on which 
$u_A^1$ generically depends linearly and homogeneously 
and the initial data $\tau^1_{A,0}, \tau^1_{A,1}$
have compact support with respect to both $p,q$ (subject to above 
vanishing conditions),
the inhomogeneous solution $\rho_A^1$ 
has non-compact support with respect to 
$p_1$. 
The solution just 
becomes constant, it ``freezes'', 
with respect to $p_1$ for sufficiently large $|p_1|$ and 
has compact support with respect to $p_2,p_3,q$. Similar remarks 
hold for the inhomogeneous solution 
$\rho_C^1$ with respect to $q_1$ and the free data 
$\gamma^2,\nu,\alpha^2,\beta^1$
on which
$u_C^1$ 
depends linearly and homogeneously and the initial data $\tau^1_{C,0},
\tau^1_{C,1}$. Identical remarks hold for $l\not=1$. This relies on the 
non-vanishing of $\vec{d}_A$ and/or $\vec{d}_C$ at generic points as argued 
in section \ref{s1.4a} which excludes the possibility that these vector 
fields have compact support.\\ 
\\
2.\\
The constraint equations at vertex $A$ and $C$ respectively only involve 
$\kappa^{1,l,\sigma}$ and $\kappa^{2,l,\sigma}$ respectively and if there 
was no constraint equation at vertex $B$ the solution of those equations 
could be found independently of each other, there would be no ``propagation''.
Assembling the various results, the generic solution found above has the 
structure ($\lambda:=-
\frac{\vec{d}_{B,1}\cdot \vec{d}_{B,2}}
{||\vec{d}_{B,1}||^4+||\vec{d}_{B,2}||^4}$)
\ba \label{1.43}
\Delta^{1,l} &=& \gamma^1\; d^l_{B,3}+
[\lambda\; ||d_{B,1}||^2\; (\alpha^2+\beta^1) +||d_{B,2}||^2 \nu]\;d^l_{B,1}
+\beta^1\; d^l_{B,2}
\nonumber\\
\Delta^{2,l} &=& \gamma^2\; d^l_{B,3}+
[\lambda\; ||d_{B,2}||^2\; (\alpha^2+\beta^1) -||d_{B,1}||^2 \nu]\;d^l_{B,2}
+\alpha^2\; d^l_{B,1}
\nonumber\\
\kappa^{1,l,-} &=&\rho_A^l=
G_A^l[\Delta^{1,l},\tau^l_{A,0},\tau^l_{A,1}]
\nonumber\\
\kappa^{2,l,-} &=&\rho_C^l=
G_C^l[\Delta^{2,l},\tau^l_{C,0},\tau^l_{C,1}]
\nonumber\\
\kappa^{I,l,+} &=& \Delta^{I,l}+\kappa^{I,l,-}
\ea
where $\alpha^{I=2},\beta^{I=1},\nu,\gamma^I;\;I=1,2$
are 5 free functions on $\mathbb{Z}^6$ and 
$\tau^l_{\ast,I},\; l=1,2,3; \ast\in \{A,C\},\;I=0,1$ 
are 12 free functions on $\mathbb{Z}^5$ (both subject to vanishing 
constraints on $\mathbb{Z}^3$ hypersurfaces). 
The functions $G_A^l,G_B^l$ are homogeneous 
linear aggregates of their arguments displayed.
Modulo characteristic crossing 
issues, this is by far not the most general solution, we expect to be 
able to generically
add homogeneous solutions to $\kappa^{I,l,-}$ which depend on 4 free 
functions on $\mathbb{Z}^6$ but this possibility will not play any 
role for what follows, because the homogeneous parts can be solved
independently at vertices A and C and thus do not contribute 
to propagation.  
 
The index $I=1$ tells that the degree of freedom 
is associated to vetex $A$ while the index $I=2$ tells that the degree of 
freedom is associated to vertex $C$. The degree of freedom $\nu$ is 
part of both $\alpha^1$ and $\beta^2$ and cannot be associated to only one 
of those vertices. Therefore $\Delta^{1,l}$ depends only on data associated 
to $A$ if we set $\nu=\alpha^2=0$ and $\Delta^{2,l}$ depends only on data 
associated to $C$ if we set $\nu=\beta^1=0$. In this case, 
$\Delta^{I,l}=\gamma^I b^l_{B,3}$ and therefore 
$\kappa^{I,l,-}$ 
depends only on 
$\gamma^I$ and
thus also $\kappa^{I,l,+}$ does. That is to say, if we set 
$\alpha^2=\beta^1=\nu=0$ and thus also $\alpha^1=\beta^2=0$ 
then the constraint equations at vertex B are trivially
satisfied and the constraint equations at vertices A and C can be solved 
independently of each other. As soon as one of $\alpha^2,\beta^1,\nu$ is 
non-vanishing, this is no longer the case. In particular, if we perturb
\footnote{This notion of `extrinsic' perturbation is distinct from the intrinsic 
disturbance/perturbation discussed in section \ref{s2.3}}
the 
solution by varying the degree of freedom $\beta^1$ associated to A 
this perturbs not only $\alpha^1$ but also $\beta^2$ associcted to $C$
unless $\vec{b}_{B,1}\cdot \vec{b}_{B,2}=0$ which will not be the case at generic values
of $p,q$.
Likewise, if we perturb
the 
solution by varying the degree of freedom $\alpha^2$ associated to C 
this perturbs not only $\beta^2$ but also $\alpha^1$ associcted to $A$.
A perturbation of $\nu$ affects both $\alpha^1,\beta^2$ and can also 
not be localised to either of the two vertices A or C.

It is straightforward to check that the solution with one of $\nu, \alpha^2, \beta^1$ non-vanishing
encodes propagation in the `intrinsic' sense of section \ref{s2.3}  and that in such  a solution there is immediate propagation from $A$ to $B$ and $C$ to $B$ but
none from $B$. The maximum propagation distance is then just unity.

Since we not only have access to  specific solutions but an entire class of 
solutions with free parameters, it is also 
possible to ask for a measure of the {\em strength} of  `extrinsic' 
propagation. A more precise definition can be found below, but 
the idea motivated by our example is as follows.
We define 
the {\it propagation degree} of a set of solutions to be the minimal 
number of its free functions that must be fixed in order that the 
parameters associated to each each vertex depends on mutually disjoint sets  
of the remaining free parameters. 
This definition makes sense because 
then the parameters for each vertex completely decouple.
We define the {\it propagation co-degree} as the set of all free parameters 
of a set of solutions minus the propagation degree. 

In our case, the parameters that 
are associated to vertex A are the $\kappa^{1,l,\sigma}$ and   
the parameters that are associated to vertex C are the 
$\kappa^{2,l,\sigma}$. By inspection of (\ref{1.43}) we see that 
in the case $\lambda=0$ (i.e. $\vec{d}_{B,1}\cdot \vec{d}_{B,2}=0$) the 
decoupling is achieved by setting $\nu=0$. In case $\lambda\not=0$
we may write 
\ba \label{1.44}
\beta^2 &:=&
\lambda\; ||d_{B,2}||^2\; (\alpha^2+\beta^1) -||d_{B,1}||^2 \nu
\nonumber\\
\alpha^1 &:=&
\lambda\; ||d_{B,1}||^2\; (\alpha^2+\beta^1) +||d_{B,2}||^2 \nu
=[\frac{||d_{B,2}||^4}{||d_{B,1}||^2}+||d_{B,1}||^2]\;\lambda\;
(\alpha^2+\beta^1)
-\frac{||d_{B,2}||^2}{||d_{B,1}||^2}\;\beta_2
\ea
and consider $\beta^2$ instead of $\nu$ as free parameter while $\alpha^1$
remains a dependent parameter. Then 
we can make $\Delta^{1,l}$ only depend on $\beta^1$ and
$\Delta^{2,l}$ only depend on $\alpha^2$ by fixing 
$\beta^2=\lambda\; \alpha^2\; 
[\frac{||d_{B,1}||^4}{||d_{B,2}||^2}+||d_{B,2}||^2]$.

We conclude that the propagation degree (to be defined below) 
of our set of solutions is 
unity. The co-degree lies between four and eight depending on 
how many free functions parametrise the set of homogeneous solutions
(neglecting the
the ``measure zero'' freedom/restictions  
provided by initial data/vanishing conditions) to the 
constraint equations. Thus at least two functions each 
can 
can be localised to the vertices A and C respectively in the sense that these  
automatically satisfy the constraint equation at B. But there is 
precisely one function that can be considered responsible 
for non-locality ($\beta^2$ in the above 
parametrisation).\\
\\
{\bf Propagation}\\
\\
The definition of  propagation in section \ref{s2.3} is based on the 
interpretation of the solution at fixed parameter values as already encoding 
a disturbance or perturbation and may be applied to any fixed  solution 
with fixed parameter values. Since this definition of propagation 
is dependent on the intrinsic structure of a single solution, 
we referred to  such propagation (see comments at the end of 
section \ref{s2.3}) as {\em intrinsic}. In contrast, we now
discuss a definition of `extrinsic' propagation based on an alternate 
`extrinsic' interpretation of 
perturbations as changes from one solution to another through small changes 
in free parameters. That `extrinsic' notion of propagation is less abstract 
and more adapted to the actual construction of solutions about which the 
`intrinsic' definition gives no information.

\begin{Definition}\label{def2.1} ~\\
1.\\
Suppose that a candidate first generation solution $\Psi$ of the quantum 
Einstein equations based on a primordial parent graph $\gamma^{(0)}$  
is a linear combination with non-redundant coefficients $\kappa_{v,i,k}$
where $v$ runs through a subset $V$ of $V(\gamma^{(0)})$ and at 
given $v\in V$, $i$ runs through an index set $I_v$ labelling 
$\gamma^{(1)}_v\in \Gamma_v(\gamma^{(0)})$ and the labels on the edges
and vertices respectively
of $\gamma_v^{(1)}$ which are not edges and vertices of $\gamma^{(0)}$
respectively. Finally $k$ runs 
through a fixed subset $L$ of the the labels of SNWF over $\gamma^{(0)}$. 
It is being understood that gauge invariance is obeyed and some of 
the $\kappa_{v,i,k}$ are automatically vanishing. It is also understood
that non-unique parentage is taken into account 
(i.e. there is no overcounting). That is, 
\be \label{2.60}  
\Psi=\sum_{v\in V}\;\sum_{i\in I_v}\;
\sum_{k\in L}\; \kappa_{v,i,k}\; \eta(T_{i,k})
\ee
Suppose that the general solution to the quantum Einstein equations imposes 
conditions on the $\kappa_{v,i,k}$ to the effect that they become 
constrained, at generic points, as follows
\be \label{2.6a}
\kappa_{v,i,k}=
\sum_{a\in A}\; \; \sum_{k'\in L}\; 
z_{(v,i),a}(k,k')\; f_a(k')
+
\sum_{b\in B}\; \; \sum_{k'\in L_b}\; 
z_{(v,i),b}(k,k')\; g_b(k')
\ee
Here $f_a$ are free functions on all of $L$ while $g_b$ are free 
functions on proper ``lower dimensional'' subsets $L_b\subset L$. As 
$L$ is discrete, lower dimensional means that $L_b$ is parametrised by less 
discrete variables than $L$ is, where those missing variables take 
infinitely many values in $L$.
Also, for each $a\in A$, $f_a$ is a linear combination of   
certain $\kappa_{v,i,l}$ for some fixed $v$ and some $i\in I_v$ which 
label the same $\gamma_a=\gamma_{v}^{(1)}\in \Gamma_v(\gamma^{(0)})$.
Let $A_v\subset A$ be such that for each $a\in A_v$ 
there exists at least one $i\in I_v$ 
such that the function $L\times L \to \mathbb{C}$ with 
$(k,k')\mapsto z_{(v,i),a}(k,k')$ is not identically zero.  
Then $\Psi$ is said 
to display propagation with respect to the choice $V$ if the $A_v$ are 
not mutually disjoint.\\
2.\\
The propagation degree $p_V$ subordinate to a choice $V$,
of a solution parametrised by the label set $A$,
is the number of elements in the smallest set $C\subset A$ such that the sets 
$A_v-C$ are mutually disjoint.  
The maximal (minimal) propagation degree $P$($p$) is the 
maximum (minimum)  of 
$p_V$ over all possible choices of $V$. The propagation co-degree 
is respectively the number of parameters in $A$ minus the respective 
propagation degree.\\
3.\\ A solution is said to be weakly propagating if $P>0$ 
(i.e. there exists $V$ 
with $p_V>0$)
and strongly propagating if $p>0$ (i.e. for all $V$ we have $p_V>0$).\\
4.\\ 
A vertex $v'\in V$ is said to be correlated with $v\in V$ if 
the sets $A_v, A_{v'}$ are not mutually disjoint.
Let $V_v\subset V$ be the subset 
of vertices correlated with $v$. 
The correlation length with respect to $V$ is the maximum of the 
set $\{d(v,v'),\; v\in V,\; v'\in V_v\}$ where $d(v,v')$ is the minimal 
number of edges in $\gamma^{(0)}$ that one needs to traverse in order 
to reach $v'$ from $v$.  
\end{Definition}

This  definition is also quite  technical, thus let 
us explain the intuitive meaning of the functions $f_a,g_b$: 
As the number of the functions $l\mapsto \kappa_{v,i,l}$ is larger 
than the number of equations (one for each in $v\in V(\gamma^{0)})$ and
$l\in L$) one will try to solve the constraint equations by separating 
the $\kappa_{v,i}$ into independent and dependent functions and solve 
the equations for the dependent functions. The $\alpha_a$ capture those 
independent functions. However, as we have seen in the context of our example,
the Hamiltonian constraint equations imposes conditions on the dependent 
$\kappa_{v,i,l}$ which can be considered as discrete partial differential 
equations (``partial difference equations''). The freedom captured by 
$\beta_b$ corresponds to the initial data on measure subsets $L_b$ 
that one needs to provide in 
a solution to those PDE's (we may equivalently define $g_b$ on all of 
$L$ by extending it trivially to $L-L_b$ in which case $g_b$ is supported 
only on measure zero subsets). The fact that we phrase propagation in terms 
of the set $A$ only rather than also $B$ is that the functions $g_b$ 
have support on ``measure zero subsets'' of $L$. If we would include     
also $B$ into the definition of propagation, this would increase the 
propagation degree, hence using only $A$ could be considered a minimal 
requirement. The physical significance of this choice is not 
clear at the moment and it may turn out in the future that this has to be 
revisited. 

Another potential issue is the dependence of the definition on
the choice of parameterization. We have reduced this choice, based on the 
structure of our example, so that each $f_a$ is constructed out of 
$\kappa_{v, i, l}$ in the manner described above. Whether this is too stringent
a requirement to be realised in generic solution constructions or not 
stringent enough
to prevent the  notion 
of propagation being dependent on the available choices of parameterization 
is also not clear at the moment. To see what the role of the $f_a$
is in our example and how they are expressed in terms of the $\kappa_{v,i,k}$
we note that in our example $V=\{A,C\}$ corresponding to $I=1,2$, 
$L=\mathbb{Z}^6$ with coordinates $k=(m_{AB}=p,m_{BC}=q)$ and 
$I_A=I_C=\{+1,-1\}\times \{1,2,3\}$ with indices $i=(\sigma,l)$. 
We confine ourselves to the generic case that the 
vectors $\vec{d}_{B,I}$ are linearly independent and set a possible 
homogeneous solution to zero. We pick as 
$f_a,\; a=\{1,2,3,4,5\}$ the functions 
$f_1=\beta^1,\; f_2=\gamma^1,\; f_3=\alpha^2,\; f_4=\beta^2,\; f_5=\gamma^2$
which allow us to write $\vec{\Delta}^1\equiv \Delta^A,\;
\vec{\Delta}^2\equiv \vec{\Delta}^B$ as linear combinations 
of the vectors $\vec{d}_{B,1},\;\vec{d}_{B,2},\;
\vec{d}_{B,3}:=\vec{d}_{B,1}\times\vec{d}_{B,2}$ (which are prescribed by 
the dynamics) as above in terms of 
$f_1,..,f_5$ while
$\alpha_1=:=[\frac{d_2^4}{d_1^2+d_1^2}]\lambda
(f_3+f_1)-\frac{d_2^2}{d_1^2}f_4$ is a dependent function, see (\ref{1.44}).
Here we have abbreviated $d_I^2:=||\vec{d}_{B,I}||^2,\;I=1,2,3$. 
Let $D$ be the 
2x2 matrix with entries $D_{IJ}:=\vec{d}_{B,I}\cdot \vec{d}_{B,J},\;
I,J=1,2$ which 
is non-degenerate as $\det(D)=||\vec{d}_{B,3}||^2$. Let also 
$\vec{\Delta}_A\cdot \vec{d}_{B,I}:=\Delta_{A,I},\;
\vec{\Delta}_C\cdot \vec{d}_{B,I}:=\Delta_{C,I},\;I=1,2,3$. Then we can write 
$f_1,..,f_5$ in terms of $\vec{\Delta}_A,\;\vec{\Delta}_C$ as follows using 
the inverse $D^{-1}$
\be \label{2.61}
f_2=\frac{\Delta_{A,3}}{d_3^2},\;
f_5=\frac{\Delta_{C,3}}{d_3^2},\;
f_1=\frac{D_{11}\;\Delta_{A,2}-D_{12}\;\Delta_{A,1}}{d_3^2},\;
f_3=\frac{D_{22}\;\Delta_{C,1}-D_{12}\;\Delta_{C,2}}{d_3^2},\;
f_4=\frac{D_{11}\;\Delta_{C,2}-D_{12}\;\Delta_{C,1}}{d_3^2}
\ee
Since with $\vec{\kappa}_A^\sigma:=\vec{\kappa}^{1,\sigma},\;
\vec{\kappa}_C^\sigma:=\vec{\kappa}^{2,\sigma}$ we have
\be \label{2.62}
\vec{\Delta}_A=\vec{\kappa}_A^+ - \vec{\kappa}_A^-,\;
\vec{\Delta}_C=\vec{\kappa}_C^+ - \vec{\kappa}_C^-
\ee
we have demonstrated that $f_1,f_2$ can be expressed as linear 
combinations of the $\kappa_A^i$ which refer to the graph 
$\gamma_A:=\gamma_1$ while $f_3,f_4,f_5$
can be expressed as linear 
combinations of the $\kappa_C^i$ which refer to the graph 
$\gamma_C:=\gamma_2$. Therefore, for the index $a=1,2$ we have the fixed 
graph $\gamma_a=\gamma_A$ associated to $v=A$ 
and for the index $a=3,4,5$ we have 
the fixed graph $\gamma_b=\gamma_C$ associated to $v=C$.  

As for the definition of intrinsic propagation, let us apply 
the above definition to our example.
Accordingly, interpreting a perturbation of a solution of its free 
coefficients associated to A that affects its free coefficients associated to 
C as extrinsic propagation of the perturbation from A to C via B, we see 
that extrinsic
propagation is a generic feature in this example class of solutions.    
The crucial feature of the system (\ref{1.30}) responsible 
for this is the non-trivial coupling
between the $\kappa^{1,l,\sigma}_M$ and $\kappa^{2,l,\sigma}_M$ coefficients
that is displayed in its third equation. The first and second condition 
respectively can be argued to provide a local condition on 
$\kappa^{1,l,\sigma}_M$ and 
$\kappa^{2,l,\sigma}_M$ respectively which solely arise due to the action of 
the Hamiltonian constraint at vertices $A$ and $C$ respectively.
However, the third equation which arises from its action at vertex $B$ 
cannot be considered as local to $B$ as it involves both 
$\kappa^{1,l,\sigma}_M,\;\kappa^{2,l,\sigma}_M$ whose ``locality
was already assigned'' to vertex A,C respectively. The reason for why this 
happens is due to the mechanism of non-unique parentage. A child vector state 
of the form $T^{l,\sigma}_{\gamma_1,M;\Delta}$ is in the image of both 
$C_A$ and $C_B$ but from different parents. The parent $T_{\gamma,m,N}$ 
for $C_A$ must adapt the charges $m_{AB},m_{BC}$ according to 
$m_{AB}=M_{AB}+\sigma \delta_l,\;m_{BC}=M_{BC}$ 
while the parent $T_{\gamma,m,N}$ for $C_B$ must adapt the charges according 
to $m_{AB}=M_{AB},\;m_{BC}=M_{BC}-\sigma\delta_l$. 
Likewise  a vector child state 
of the form $T^{l,\sigma}_{\gamma_2,M;\Delta}$ is in the image of both 
$C_C$ and $C_B$ but from different parents. The parent $T_{\gamma,m,N}$ 
for $C_C$ must adapt the charges $m_{AB},m_{BC}$ according to 
$m_{AB}=M_{AB},\;m_{BC}=M_{BC}-\sigma\delta_l$ 
while the parent $T_{\gamma,m,N}$ for $C_B$ must adapt the charges according 
to $m_{AB}=M_{AB}+\sigma\delta_l,\;m_{BC}=M_{BC}$. This means that the 
coefficients 
{\it cannot be unambiguously associated to vertices} in the sense that 
one has disjoint sets of coefficients that are to be solved for 
vertex wise with no further conditions. \\
\\
We conclude that the example studied displays intrinsic propagation 
in the sense of section \ref{s2.3} 
(i.e. in the context of a fixed solution) as well as in the extrinsic sense
above in terms of perturbations of free parameters which 
define  the space of solutions.

\section{Conclusion and outlook}
\label{s3}

It is obvious that the propagation effect sketched above 
extends to more complicated graphs (e.g. those that are 
networks made out of arbitrarily long and knotted chains 
of four valent graphs intersecting in eight valent vertices and possibly
with some knotted loops attached to make the graph symmetry group trivial)
and will even be enhanced
the higher the connectivity of the graphs and the valence 
of their vertices are, at least in the U(1)$^3$ 
theory. That is, we expect that the propagation degree of a solution
class, as well as the propagation distance in a (propagating) solution, that is labelled by more and more complicated graphs will drastically
increase with the number and complexity of the graphs considered.   

As far as the extension to SU(2) is concerned, note that the qualitative 
features of our example remain the the same. The curvature 
factor in the Hamiltonian constraint now leads to shifts by $\pm \frac{1}{2}$ 
in the spin quantum numbers on the edges of  
SNWF and the arcs get charged with spin $1/2$ rather than 
$\pm \delta_l$. The only difference is that in the SU(2) case 
we are not able to compute the eigenvalues of the volume 
operator so easily and this was used above to give an easy proof of the 
absence of the disjointness of sets of coefficients required for 
presence of propagation 
within a solution. However, it is hard to imagine that there should 
be an accidental symmetry that would render magically all those matrix 
elements of the volume operator to zero which would lead to 
absence of long range 
correlations of the solution of the Hamiltonian constraint. At least,
the present paper establishes that the burden of proof rests 
on those that argue that the locality of the action of the 
Hamiltonian constraint at vertices implies locality of its solution with no 
propagation. In fact, our paper suggests a numerical proof 
of propagation both for Lorentzian and 
Euclidean GR with gauge group SU(2) by diagonalising 
the volume operator on four valent vertices 
numerically 
which is 
feasable since the matrix elements of its fourth 
power are available analytically \cite{JBTT}. 
We leave this to future work which may benefit from modern 
numerical \cite{QGC,Num} methods and machine learning \cite{AI,TN} and/or 
quantum computing \cite{QC} techniques.

Finally, in our example we considered only a tiny subset 
of solutions as we took only children of ``first generation'' into account.
More generally one can consider solutions that involve the images 
of an arbitrary number of applications of the Hamiltonian constraint 
on CNWF. In the Euclidean theory, the set of conditions to be met 
by a solution to the constraints does not mix those generations and in 
that sense the analysis is complete when one controls the one generation 
case. 
However, this no longer true in the Lorentzian theory which does 
mix generations.
Hence we expect an even stronger notion of propagation
for Lorentzian signature. See \cite{QSDII} for details.

We close by stressing again that propagation hinges on the presence of 
non-unique parentage. This is the case, modulo the 
above reservations, in QSD \cite{QSDI,QSDII} and 
in the electric shift approach of \cite{MV} even for the Euclidean 
part of the constraint but generically 
not e.g. for the symmetric Euclidean  
Hamiltonian constraint operator proposal of \cite{LS}: There the 
the loops created by the Hamiltonian constraint are attached only to 
the vertex of the original graph, i.e. they intersect the original graph
in that single vertex and no other point of the graph. Moreover, they
lie in the coordinate plane defined by the tangents of two edges adjacent
to that vertex. Accordingly, unless there are graph symmetries,
such a child graph has a unique parent graph even after diffeomorphism 
averaging. Therefore constraint equations at each vertex are local to 
that vertex. As the loop attachment does not change the spins of the edges
of the parent SNWF, the Euclidean 
Hamiltonian constraint at a given vertex 
thus only imposes constraints on the intertwiners associated to that vertex
and that intertwiner space is finite dimensional. Thus the solutions 
to the Euclidean Hamiltonian of \cite{LS} are normalisable 
with respect to the norm of the diffeomorphism invariant Hilbert space 
in contrast to the situation in \cite{QSDI} and the present paper.

\begin{appendix}

\section{\label{seca1}Explicit form of $d^l_A,d^l_C,d^l_{B,1},d^l_{B,2} $}

From (\ref{1.6}) and (\ref{1.7}) we have that:
\be
Q_A = (m_{ZA} +c_A,\; n_{ZA},\; m_{AB}- n_{AB}- c^{\prime}_A)  - (m_{AB},\; n_{AB}+c^{\prime}_A,\; m_{ZA} +c_A-n_{ZA})
\ee
\be
d^l_A := -d^l_{A,\;1}  +d^l_{A,\;2}  +d^l_{A,\;3} -  d^l_{A,\;4}  
\ee
with 
\ba
d^l_{A,\;1}  &=& |(Q_A +( n_{AB}+c_A,\;n_{ZA}+m_{AB},\;  \delta_l) - (m_{AB},\;n_{ZA},\; \delta_l))|^{\frac{1}{2}}
\\
d^l_{A,\;2}  &=& |(Q_A - ( n_{AB}+c_A,\;n_{ZA}+ m_{AB},\;   \delta_l) + (m_{AB},\;n_{ZA},\; \delta_l))|^{\frac{1}{2}}
\\
d^l_{A,\;3}  &=& |(Q_A - ( n_{AB}+c_A,\; m_{ZA} +c_A +m_{AB},\; \delta_l) + (m_{AB} ,\;m_{ZA}+ c_A,\;\delta_l))|^{\frac{1}{2}}
\\
d^l_{A,\;4}  &=& |Q_A |^{\frac{1}{2}}
\ea
where we have found it convenient to denote the $\nu$ functions in (\ref{1.7}) by $d^l_{A,i}$. We shall continue to use such notation
in what follows.

Next, from (\ref{1.8}) and (\ref{1.9}) we have that:
\be
Q_C = (m_{BC} +c_C,\; n_{BC},\; m_{CD}- n_{CD}- c^{\prime}_C)  - (m_{CD},\; n_{CD}+c^{\prime}_C,\; m_{BC} +c_C-n_{BC})
\ee
\be
d^l_C := d^l_{C,\;1}  -d^l_{C,\;2}  -d^l_{C,\;3} +  d^l_{C,\;4}  
\ee
with 
\ba
d^l_{C,\;1}  &=& |(Q_C +( m_{BC}+c_C,\;n_{BC}+m_{CD},\;  \delta_l) - (n_{BC},\;m_{CD},\; \delta_l))|^{\frac{1}{2}}
\\
d^l_{C,\;2}  &=& |(Q_C -( m_{BC}+c_C,\;n_{BC}+m_{CD},\;  \delta_l) + (n_{BC},\;m_{CD},\; \delta_l))|^{\frac{1}{2}}
\\
d^l_{C,\;3}  &=& |(Q_C - ( m_{BC}+c_C,\; n_{BC} +c_C +n_{CD},\; \delta_l) + (n_{BC} ,\;n_{CD}+ c_C,\;\delta_l))|^{\frac{1}{2}}
\\
d^l_{C,\;4}  &=& |(Q_C |^{\frac{1}{2}}
\ea

Finally, from (\ref{1.11}), (\ref{1.12}) and (\ref{1.13}) we have:
\be
Q_B = (m_{AB} +c_B,\; n_{AB},\; m_{BC}- n_{BC}- c^{\prime}_B)  - (m_{BC},\; n_{BC}+c^{\prime}_B,\; m_{AB} +c_B-n_{AB})
\ee
\be
d^l_{B,1} := d^l_{B1,\;1}  -d^l_{B1,\;2}  -d^l_{B1,\;3} +  d^l_{B1,\;4}  
\ee
with 
\ba
d^l_{B1,\;1}  &=& |(Q_B +( m_{AB}+c_B,\;n_{AB}+m_{BC},\;  \delta_l) - (n_{AB},\;m_{BC},\; \delta_l))|^{\frac{1}{2}}
\\
d^l_{B1,\;2}  &=& |(Q_B -( m_{AB}+c_B,\;n_{AB}+m_{BC},\;  \delta_l) + (n_{AB},\;m_{BC},\; \delta_l))|^{\frac{1}{2}}
\\
d^l_{B1,\;3}  &=& |(Q_B - ( m_{AB}+c_B,\; n_{AB} + n_{BC}+c_B,\; \delta_l) + (n_{AB} ,\;n_{BC}+ c_B,\;\delta_l))|^{\frac{1}{2}}
\\
d^l_{B1,\;4}  &=& |(Q_B |^{\frac{1}{2}}
\ea
\be
d^l_{B,2} := d^l_{B2,\;1}  -d^l_{B2,\;2}  -d^l_{B2,\;3} +  d^l_{B2,\;4}  
\ee
with 
\ba
d^l_{B2,\;1}  &=& |(Q_B - (m_{BC}+n_{AB},\;n_{BC}+c_{B},\;  \delta_l) + (n_{AB},\;m_{BC},\; \delta_l))|^{\frac{1}{2}}
\\
d^l_{B2,\;2}  &=& |(Q_B -( m_{AB}+c_B,\;m_{BC},\;  \delta_l) + (m_{AB}+c_B+m_{BC},\;n_{BC}+c_B,\; \delta_l))|^{\frac{1}{2}}
\\
d^l_{B2,\;3}  &=& |(Q_B +( m_{BC}+n_{AB},\;n_{BC}+c_{B},\;  \delta_l) - (n_{AB},\;m_{BC},\; \delta_l))|^{\frac{1}{2}}
\\
d^l_{B2,\;4}  &=& |Q_B |^{\frac{1}{2}}
\ea

\end{appendix}

\end{document}